\documentclass[twocolumn]{aastex6}

\usepackage{amsmath}
\usepackage{mathtools}
\usepackage{amssymb} 
\usepackage{bm}
\usepackage{ulem}

\usepackage{lipsum}
 
\shorttitle{Net Radial Flow in NGC 4736}
\shortauthors{Speights et al.}

\begin{document}

\title{The Net Radial Flow Velocity of the Neutral Hydrogen\\ in the Oval Distortion of NGC 4736}

\author{Jason C. Speights, Caleb Godwin, Rebecca Reimer, Allen Benton, \& Robert Lemaire}
\affil{Frostburg State University\\
101 Braddock Rd. \\
Frostburg, MD 21532, USA}

\altaffiltext{1}{jcspeights@frostburg.edu}

\begin{abstract}

The net radial flow velocity of gas is an important parameter for understanding galaxy evolution. It is difficult to measure in the presence of the elliptical orbits of an oval distortion because the mathematical model describing the observed velocity is degenerate in the unknown velocity components. A method is developed in this paper that breaks the degeneracy using additional information about the angular frequency of the oval distortion. The method is applied to the neutral hydrogen in the oval distortion of NGC 4736. The neutral hydrogen is flowing inward at a mean rate of -6.1 $\pm$ 1.9 km s$^{-1}$. At this rate, it takes 400 Myr, or 1.7 rotations of the oval distortion, for the neutral hydrogen to travel the 2.5 kpc from the end to the beginning of the oval distortion. The mean mass flow rate of the neutral hydrogen in this region is -0.25 $\pm$ 0.11 $M_\odot$ yr$^{-1}$, which is similar to estimates for the star formation rate reported in the literature. 

\end{abstract}


\section{Introduction} \label{sec:intro}

As redshift decreases from the first few billion years of the universe to its current age, major mergers and interactions occur less frequently (Toomre 1977, Conselice et al. 2003, lavery et al. 2004, de Ravel et al. 2009, Bridge et al. 2010, Mantha et al. 2018) and internal, secular processes play an increasingly larger role in galaxy evolution (Kormendy \& Kennicutt 2004). Oval distortions, like bars, are well-known engines for secular evolution owing to their ability to torque the gas and drive it radially inward  (Fukunaga \& Tosa 1991; Friedli \& Benz 1993; Shlosman \& Noguchi 1993; Berentzen et al. 1998; Sakamoto et al. 1999; Combes 2008; Fanali et al. 2015). The purpose of this paper is to develop and apply a method for measuring the net radial flow velocity of neutral hydrogen (H{\footnotesize I}) in an oval distortion. 

The expected values for the the net radial flow velocities vary from as small as $\approx$ -0.1 km s$^{-1}$ to as large as $\approx$ -10 km s$^{-1}$. (e.g., Lacey \& Fall 1985; Struck-Marcell 1991; Athanassoula 1992; Quillen et al. 1995; Struck \& Smith 1999; Bilitewski \& Sch{\"o}nrich 2012; Schmidt et al. 2016).  Although the expected net radial flow velocities are typically 1--2 orders of magnitude smaller than the circular velocities, the inflowing gas can greatly affect galaxy evolution. A velocity of $\approx$ 1 km s$^{-1}$ converts to $\approx$1 kpc Gyr$^{-1}$.

Inward gas flows are needed to replenish the fuel for star formation in the central regions of galaxy disks (Bigiel at al. 2011; Rahman et al. 2012; Utomo et al. 2017). There, it can lead to star forming rings (Athanassoula 1983; Buta \& Combes 1996; Combes 1996; Jungwiert \& Palous 1996; Kim et al. 2014; Li et al. 2015), enhanced star formation rates (Jogee et al. 2005; Ellison et al. 2011; Lin et al. 2017), and excess central gas concentrations (Kenney et al. 1992; Regan et al. 2001; Sheth et al. 2005). For some galaxies, the gas finds a way to the very center and feeds active galactic nuclei (AGNs, e.g., Shlosman et al. 1989).

These flows shape a galaxy's morphology (e.g., Lin \& Pringle 1987) and ultimately contribute to the formation of pseudobulges (Courteau et al. 1996; Jogee et al. 2005; Fisher et al. 2013). The increasing mass density of a growing pseudobulge can weaken and dissolve the bar or oval that built it (Hasan \& Norman 1990; Pfenniger \& Norman 1990; Norman et al. 1996; Bournard \& Combes 2004). Some oval distortions may be the remnants of dissolved bars (Kormendy 1979; Berentzen et al. 2006; Laurikainen et al. 2009).

\begin{figure*}[ht!]
\centering
\includegraphics[width=1\textwidth]{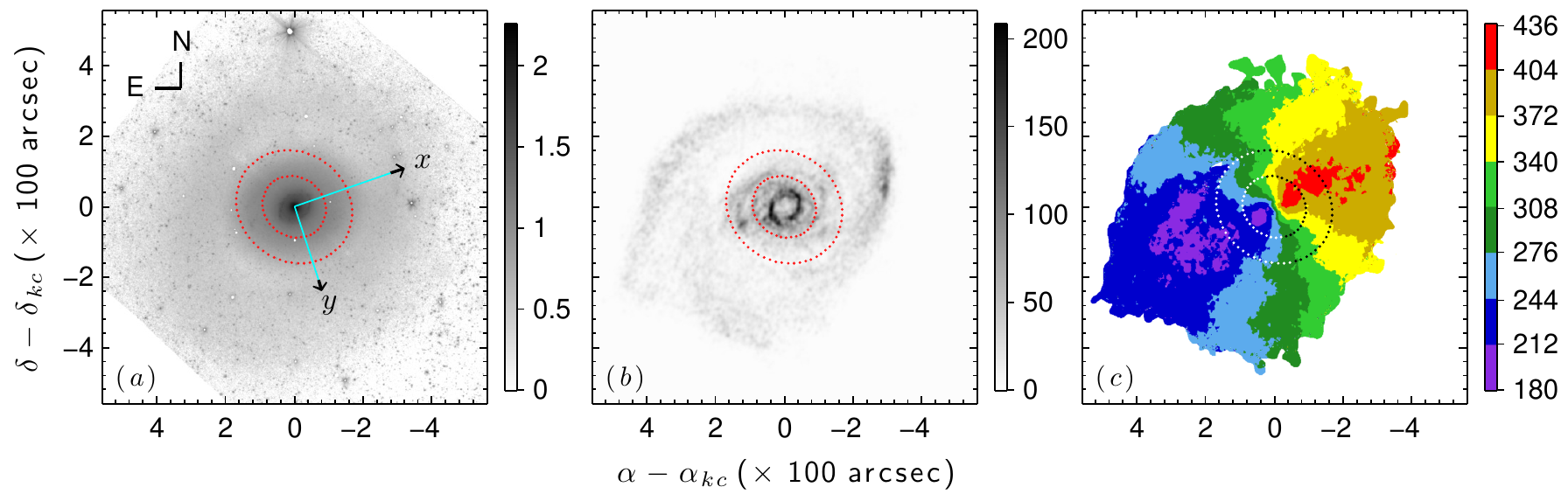} 
\caption{NIR and H{\footnotesize I} data. Panels ($a$)--($c$) show $I_{3.6}$ for the NIR data, $I_{HI}$ for the H{\footnotesize I} data, and  $V_{\mbox{\footnotesize los}}$ for the H{\footnotesize I} data, respectively. The color bars are in units of log(MJy sr$^{-1}$) for panel ($a$), Jy beam$^{-1}$ m s$^{-1}$ for panel ($b$), and km s$^{-1}$ for panel ($c$). The directions for N and E in the sky, and $x$ and $y$ in the galaxy coordinates (Section 3.1), are shown in panel ($a$). Most of the foreground starlight from the Milky Way is removed from panel ($a$) using Source Extractor (Bertin \& Arnouts 1996, hereafter BA96). In panel ($c$) the data are binned in increments of 32 km s$^{-1}$. The black arrows in panel ($a$) that are at the end of the cyan lines for $x$ and $y$ are 1 kpc long in the plane of the galaxy disk, assuming a distance of 5.1 Mpc, and point to $R_{\mbox{\footnotesize 25}}$ (Section 3.2). The red dotted lines in panels ($a$) and ($b$) trace the inner and outer radii of the oval distortion (Sections 4.2 and 5.3.2). For clarity, the same dotted lines in panel ($c$) are colored black and white for the receding and approaching sides of the galaxy, respectively.} 
\end{figure*}

Despite its importance, there is little progress to directly measure the net radial flow velocity for the gas in an oval distortion or bar. Inward flow velocities of $\sim$ 100 km s$^{-1}$ are found along dust lanes in strong bars, but these must be interpreted as upper limits for the net radial flow velocity owing to the complex geometry of the gas orbits (Kenney 1994; Benedict et al. 1996; Regan et al. 1997, 1999; Schinnerer et al. 2002). Mathematical models for elliptical orbits are less complex (e.g., Spekkens \& Sellwood 2007, hereafter SS07; Sellwood \& S{\'a}nchez 2010, hereafter SS10), but additional information is needed for including the net radial flow velocity. A model with both elliptical orbits and net radial flows is degenerate in the unknown velocities (e.g., Wong et al. 2004; Haan et al. 2009). The method developed in this paper breaks the degeneracy using additional information about the angular frequency of the oval distortion, $\Omega_o$. 

The method is applied to the oval distortion of NGC 4736. This galaxy is one of the best known candidates for secular evolution driven by an oval distortion (Kormendy \& Kennicutt 2004, and references therein). Its morphological classification is (R)SAB(rs)ab (de Vaucoulers al. 1991, hereafter dV91). The oval distortion extends from $r$ $\approx$ 120$^{\prime\prime}$ to $r$ $\approx$ 220$^{\prime\prime}$ (M{\"o}llenhoff et al. 1995, hereafter M95). The observed velocity field of the oval distortion is inadequately described by elliptical orbits alone and shows evidence for net radial flows (Wong \& Blitz 2000, hereafter WB00). NGC 4736 has both outer and inner pseudorings of tightly wound spiral patterns. The spiral nature of the outer pseudoring centered at $\approx$ 350$^{\prime\prime}$ is visible in optical images using very long exposure times (Trujillo et al. 2009, hereafter T09). The inner pseudoring centered at $\approx$ 45$^{\prime\prime}$ is undergoing intense star formation (Buta 1988; van der Laan et al. 2015). The central region contains a nuclear bar of $\approx$ 25$^{\prime\prime}$ in radius (M95) and a low-luminosity AGN (K{\"o}rding 2005; Maoz 2005; Constantin 2012; van Oers 2017). For an assumed distance of 5.1 Mpc from the mean of the estimates reported in the NASA/IPAC Extragalactic Database, 1$^{\prime\prime}$ in the sky converts to 24.7 pc at NGC 4736, and 1 kpc at NGC 4736 converts to 40\farcs4 in the sky.

The rest of this paper is organized as follows. Section 2 describes the data. Section 3 explains the mathematical models. Section 4 presents the results. Section 5 discusses the results. Section 6 is a summary.

\section{Data} \label{sec:dat}

The mathematical models are fit to two different types of data. Near-infrared (NIR) data are used for a model that measures the phase angle, $\theta_o$, for the location of the major axis of the oval distortion. The NIR data are an image of the 3.6 $\mu$m intensity, $I_{3.6}$, by Kennicutt et al. (2003). Neutral hydrogen data are used for a model that measures $\Omega_o$, and models that measure the velocity components of the gas. The H{\footnotesize I} data are the integrated intensity, $I_{HI}$, and intensity-weighted line-of-sight velocity, $V_{\mbox{\footnotesize los}}$, for the 21 cm spectral line. These are from naturally weighted moment 0 and moment 1 maps, respectively, by Walter et al. (2008, hereafter W08). 

The data are shown in Figure 1. Comparing $I_{3.6}$ and $I_{HI}$ for the different data reveals similarities and differences in the patterns they trace. Both show the pseudoring in the nuclear region. The H{\footnotesize I} data excel at showing the spiral pattern in the outer disk and show short spiral arcs in the region of the oval distortion. The NIR data excel at showing the oval distortion and the nuclear bar. The orientation of the oval distortion is shown more clearly in Section 4.2, and discussed in Section 5.3.2.

The assumed orientation of NGC 4736 follows from the direction of the spiral patterns in the $I_{3.6}$ and $I_{HI}$ data, as well as the receding and approaching sides of the galaxy in the $V_{\mbox{\footnotesize los}}$ data. The direction of the spiral patterns is the same for the inner and outer pseudorings. The spiral patterns are assumed to be trailing, so according to the $V_{\mbox{\footnotesize los}}$ data the northeast part of the galaxy is tilted toward the observer and the southwest is tilted away. 

The properties of the data that are needed for fitting the mathematical models are the pixel size and the spatial resolution. The pixel size is 0\farcs75 for the NIR data and 1\farcs5 for the H{\footnotesize I} data.  The adopted resolution of the NIR data is the FWHM of the point-spread function, which is 1\farcs4. The adopted resolution of the H{\footnotesize I} data is the FWHM of the synthesized beam, which has a major axis of 10\farcs22, a minor axis of 9\farcs07, and a position angle of 337$^\circ$.  It is assumed that pixels within the spatial resolution of one another are correlated. 

\section{Mathematical Models} \label{subsec:meth_math}

\subsection{Coordinate System Definitions}

Conventions for relating galaxy coordinates to those observed in the sky are well rehearsed in the literature (e.g., Van der Kruit \& Allen 1978, hereafter VdKA78). This first subsection provides a brief review and defines the notation in the rest of this paper. 

The galaxy is observed in the sky at a kinematic center in R.A., $\alpha_{\mbox{\footnotesize kc}}$, and in decl., $\delta_{\mbox{\footnotesize kc}}$. It is inclined at an angle, $\psi_{\mbox{\footnotesize inc}}$, defined so that $\psi_{\mbox{\footnotesize inc}}$ = 0$^\circ$ for face-on. The receding side of the kinematic major axis is at a position angle, $\phi_{\mbox{\footnotesize maj}}$, measured from north to east. 

Cartesian coordinates in the galaxy are defined so that the positive $x$-axis is along the receding side of the kinematic major axis.  Distances measured in the $x$-direction are related to R.A., $\alpha$, and decl., $\delta$, according to 
\begin{equation}
x = (\alpha - \alpha_{\mbox{\footnotesize kc}})\,\mbox{sin}(\phi_{\mbox{\footnotesize maj}}) + (\delta - \delta_{\mbox{\footnotesize kc}})\,\mbox{cos}(\phi_{\mbox{\footnotesize maj}}).
\end{equation}
Similarly, for distances measured in the $y$-direction,
\begin{equation}
y = \frac{-(\alpha - \alpha_{\mbox{\footnotesize kc}})\,\mbox{cos}(\phi_{\mbox{\footnotesize maj}}) + (\delta - \delta_{\mbox{\footnotesize kc}})\,\mbox{sin}(\phi_{\mbox{\footnotesize maj}})}{\mbox{cos}(\psi_{\mbox{\footnotesize inc}})}.
\end{equation}
The positive and negative signs on the right-hand side of Equation (2) are chosen so that the observed net radial flow velocities are positive for outflows and negative for inflows given the assumed orientation of NGC 4736. For cylindrical coordinates in the galaxy disk, positive $\theta$ is defined from $x$ to $y$, with $\theta$ = 0$^\circ$ along the positive $x$-axis. 

\subsection{Models for the Velocities}

The model for measuring net radial flow velocities assumes that the dominating components of the H{\footnotesize I} velocity are elliptical orbits and net radial flows.  Mathematically, the $V_{\mbox{\footnotesize los}}$ data at radius $r$ have the form
\begin{align}
V_{\mbox{\footnotesize los}}(r,\theta) = \hskip 2pt & V_{\mbox{\footnotesize sys}} + \mbox{sin}(\psi_{\mbox{\footnotesize inc}})\{V_{\theta \hskip 0.5pt 0}(r)\,\mbox{cos}(\theta)+V_{{r} \hskip 0.5pt 0}(r)\,\mbox{sin}(\theta) \nonumber \\ & - V_{\theta \hskip 0.5pt 2}(r)\,\mbox{cos}(\Theta)\,\mbox{cos}(\theta)  \nonumber \\ & - V_{{r} \hskip 0.5pt 2}(r)\,\mbox{sin}(\Theta)\,\mbox{sin}(\theta)\},
\end{align}
where that $V_{\mbox{\footnotesize sys}}$ is the systemic velocity, $V_{\theta \hskip 0.5pt 0}$ is the azimuthal (circular) velocity, $V_{r \hskip 0.5pt 0}$ is the net radial flow velocity, $V_{\theta \hskip 0.5pt 2}$ is the amplitude of the velocity perturbation in the azimuthal direction induced by the elliptical orbit, and $V_{r \hskip 0.5pt 2}$ is the  amplitude of the velocity perturbation in the radial direction  induced by the elliptical orbit. The substitution, $\Theta$ = $2\,(\theta - \theta_o)$, shortens the notation for including $\theta_o$. Equation (3) is hereafter referred to as the full model (FM).

The third and fourth terms in the curly brackets on the right-hand side of the FM modify a circular orbit to make it elliptical. The third term describes how material slows down in the azimuthal direction while approaching the major axis of the ellipse and speeds up while approaching the minor axis. The fourth term describes how material moves outward in the radial direction while approaching the major axis of the ellipse and moves inward while approaching the minor axis.

The elliptical orbits described by the FM assume $x_1$-type orbits aligned parallel to the oval distortion (Contopoulos 1980, hereafter C80). These are expected to occur between an inner Lindblad resonance and a corotation resonance. The results for the locations of resonances in Section 4.4 are consistent with this assumption. NGC 4736 may also contain $x_2$-type orbits, which are perpendicular to $x_1$-type orbits if they are rotating at the same angular frequency. These are expected to occur interior to an inner Lindblad resonance. For a more detailed discussion of orbit families and resonances, the interested reader is referred to Sellwood \& Wilkinson (1993, hereafter SW93).

The FM is a modified version of the one used by SS07 and SS10 for elliptical orbits. The difference is the inclusion of $V_{{r} \hskip 0.5pt 0}$ in this paper. They justify excluding $V_{{r} \hskip 0.5pt 0}$ by assuming that the values are too small to significantly affect the results; otherwise, there is a continuity problem. The results from assuming $V_{{r} \hskip 0.5pt 0}$ = 0 and the continuity problem are discussed in Section 5.4.

A unique fit for all of the variables in the FM is impossible for two reasons. The first is that it is nonlinear in all of the variables except $V_{\mbox{\footnotesize sys}}$. The second is that even if assumptions, other models, and more data can remove the nonlinearity, the system of equations for the remaining velocity variables are rank deficient, i.e., there is a degeneracy problem. The rest of this subsection discusses these issues and explains a method to resolve them.

Some of the nonlinearity is removed by adopting the mean values of $\alpha_{\mbox{\footnotesize kc}}$ and $\delta_{\mbox{\footnotesize kc}}$ found by de Blok et al. (2008, hereafter dB08). Their results are preferred to other estimates because their data are from the same observations W08 used for making the H{\footnotesize I} data in this paper. The mean value of $V_{\mbox{\footnotesize sys}}$ found by dB08 is also adopted. Although it does not contribute to the issue of nonlinearity, adopting it helps simplify the fitted model.

Unlike their results for $\alpha_{\mbox{\footnotesize kc}}$, $\delta_{\mbox{\footnotesize kc}}$, and $V_{\mbox{\footnotesize sys}}$, the results dB08 find for $\psi_{\mbox{\footnotesize inc}}$ and $\phi_{\mbox{\footnotesize maj}}$ are poorly described by mean values. Their model of the observed velocity allows $\psi_{\mbox{\footnotesize inc}}$ and $\phi_{\mbox{\footnotesize maj}}$ to vary in successive annuli, or rings of data, a method commonly referred to as tilted rings. Inside R$_{\mbox{\footnotesize 25}}$ = 336\farcs6 (dV91), $\psi_{\mbox{\footnotesize inc}}$ varies by $\approx$34$^\circ$, and $\phi_{\mbox{\footnotesize maj}}$ varies by $\approx$25$^\circ$. A similar result is found by Mulder \& Van Driel (1993, hereafter MVD93) and WB00. It is unlikely that these large variations are from a warp in the disk. Warps are rare within $R_{\mbox{\footnotesize 25}}$  (Briggs 1990). 

The large variations are a result of the uniqueness problem. The variation in $\phi_{\mbox{\footnotesize maj}}$, for example, is partly a result of variation in the unaccounted-for radial velocity components in galaxy coordinates (e.g., Warner et al. 1973, vdAK78).  The FM shows that an incorrect measure of $\psi_{\mbox{\footnotesize inc}}$ in sin($\psi_{\mbox{\footnotesize inc}}$) can be accounted for with incorrect measures of the velocity components in the curly brackets. If $\psi_{\mbox{\footnotesize inc}}$ were larger, one could more easily distinguish variation in $\phi_{\mbox{\footnotesize maj}}$ from the effect of radial velocities in regions with only circular and net radial velocities because the net radial velocities rotate the kinematic major and minor axes by different amounts in sky coordinates.

To measure $\psi_{\mbox{\footnotesize inc}}$ and $\phi_{\mbox{\footnotesize maj}}$, it is noted that in dB08 $\phi_{\mbox{\footnotesize maj}}$ is approximately constant between the end of the oval at  220$^{\prime\prime}$ and R$_{\mbox{\footnotesize 25}}$. This is also demonstrated in Section 4.1. The reduced model,
\begin{equation}
V_{\mbox{\footnotesize los}}(r,\theta)- V_{\mbox{\footnotesize sys}} = \mbox{sin}(\psi_{\mbox{\footnotesize inc}})\,V_{\theta \hskip 0.5pt 0}(r)\,\mbox{cos}(\theta),
\end{equation}
is fit to this region using a range of values for $\psi_{\mbox{\footnotesize inc}}$ and $\phi_{\mbox{\footnotesize maj}}$. A weighting function of $|$cos($\theta$)$|$ is used for giving greater weight to the data near the kinematic major axis. The values of $\psi_{\mbox{\footnotesize inc}}$ and $\phi_{\mbox{\footnotesize maj}}$ for the fit with the smallest sum of the squared residuals (SSRs) are adopted for the rest of this paper. 

The remaining variable that is needed for making the FM linear in the unknowns is $\theta_o$. This is found by assuming that it is coincident with the major axis of the oval distortion traced by the NIR data. The mathematical model for finding $\theta_o$ is explained in Section 3.3.

With the adopted and measured variables in the preceding paragraphs, the FM simplifies to 
\begin{align}
V_y(r,\theta) = \hskip 2pt & V_{\theta \hskip 0.5pt 0}(r)\,\mbox{cos}(\theta)+V_{{r} \hskip 0.5pt 0}(r)\,\mbox{sin}(\theta) \nonumber \\ &  - V_{\theta \hskip 0.5pt 2}(r)\,\mbox{cos}(\Theta)\,\mbox{cos}(\theta)  \nonumber \\ &- V_{{r} \hskip 0.5pt 2}(r)\,\mbox{sin}(\Theta)\,\mbox{sin}(\theta),
\end{align}
where,
\begin{equation}
V_y(r,\theta) = \frac{V_{\mbox{\footnotesize los}}(r,\theta) - V_{\mbox{\footnotesize sys}}}{\mbox{sin}(\psi_{\mbox{\footnotesize inc}})}.
\end{equation}
Equation (5) is linear in $V_{\theta \hskip 0.5pt 0}$, $V_{{r} \hskip 0.5pt 0}$, $V_{\theta \hskip 0.5pt 2}$, and $V_{r \hskip 0.5pt 2}$. The matrix of the remaining independent variables,
\begin{eqnarray}\nonumber
\left(
\begin{array}{l l l l} 
\mbox{cos}(\theta_1) & \hskip 2pt \mbox{sin}(\theta_1) & \hskip 2pt - \mbox{cos}(\Theta_1)\,\mbox{cos}(\theta_1) & \hskip 2pt - \mbox{sin}(\Theta_1)\,\mbox{sin}(\theta_1) \\
\mbox{cos}(\theta_2) & \hskip 2pt \mbox{sin}(\theta_2) & \hskip 2pt - \mbox{cos}(\Theta_2)\,\mbox{cos}(\theta_2) & \hskip 2pt - \mbox{sin}(\Theta_2)\,\mbox{sin}(\theta_2) \\
\mbox{cos}(\theta_3) & \hskip 2pt \mbox{sin}(\theta_3) & \hskip 2pt - \mbox{cos}(\Theta_3)\,\mbox{cos}(\theta_3) & \hskip 2pt - \mbox{sin}(\Theta_3)\,\mbox{sin}(\theta_3) \\
\mbox{cos}(\theta_4) & \hskip 2pt \mbox{sin}(\theta_4) & \hskip 2pt - \mbox{cos}(\Theta_4)\,\mbox{cos}(\theta_4) & \hskip 2pt - \mbox{sin}(\Theta_4)\,\mbox{sin}(\theta_4) \\
\multicolumn{1}{c}{\vdots} &\multicolumn{1}{c}{\vdots} & \multicolumn{1}{c}{\vdots} &\multicolumn{1}{c}{\vdots} \\
\mbox{cos}(\theta_n) & \hskip 2pt \mbox{sin}(\theta_n) & \hskip 2pt - \mbox{cos}(\Theta_n)\,\mbox{cos}(\theta_n) & \hskip 2pt - \mbox{sin}(\Theta_n)\,\mbox{sin}(\theta_n) 
\end{array}
\right),
\end{eqnarray}\\
for a system of $n$ linear equations for $n$ values of $V_y$ has only three linearly independent columns. This is easily demonstrated using a rank-finding function in software such as NumPy (van der Walt et al. 2011) or MATLAB\footnote{The Mathworks, Inc., Natick, Massachusetts, United States, http://www.mathworks.com}.  There is only enough information for finding a unique fit for three of the remaining variables. Additional information is required for uniquely finding all four of them. 

The degeneracy is breakable using an additional model for measuring the angular frequency, $\Omega_o$, of the oval distortion.  It is assumed that there is a single oval distortion from $r$ $\approx$ 120$^{\prime\prime}$ to $r$ $\approx$ 220$^{\prime\prime}$ rotating at a constant $\Omega_o$. Although it is possible that there is more than one oval distortion, for example, one weak and another strong, that could produce very different velocity perturbations than those assumed in the FM, the results for $\theta_o$ in Section 4.2 are consistent with a single oval distortion. The generalized form of the Tremaine \& Weinberg (1984, hereafter TW84) method applied to the H{\footnotesize I} data is used for measuring $\Omega_o$. The TW84 method and its generalized form are explained in Section 3.4. 

The motivation for using $\Omega_o$ to break the degeneracy is based on observation. Speights \& Rooke (2016, hereafter SR16) show that most of the residual velocities, $V_{res}$, from fitting a model of only circular velocities to the region of the bar in NGC 1365 are explained by the velocity perturbations induced by the bar rotating at an angular frequency of $\Omega_b$. They fit a model of the form
\begin{align}
V_{res}(r,\theta) = & - V_{\theta \hskip 0.5pt 2}(r)\,\mbox{cos}(2[\theta - \theta_b])\,\mbox{cos}(\theta)\nonumber \\  & - V_{{r} \hskip 0.5pt 2}(r)\,\mbox{sin}(2[\theta - \theta_b])\,\mbox{sin}(\theta)
\end{align}
to the residuals in the region of the bar, where that $\theta_b$ is the location of the bar major axis. Wade \& Speights (2018) fit Equation (5) to the region of the bar in NGC 4321 without the second term on the right-hand side, and with $\theta_o$ replaced by $\theta_b$. Both authors then estimate $\Omega_b$ at different radii by calculating
\begin{equation}
\Omega_b^\prime(r) = \frac{V_{\theta \hskip 0.5pt 0}(r) - V_{\theta \hskip 0.5pt 2}(r)}{r},
\end{equation}
using the results for $V_{\theta \hskip 0.5pt 0}$ and $V_{\theta \hskip 0.5pt 2}$ from the fitted models. Their results are found to be consistent with the results from using the TW84 method. Note that prime notation is used in this paper to differentiate from the results obtained using the TW84 method.

The same relationship is found to hold for NGC 4736. In Section 4.4 it is shown that the results from calculating
\begin{equation}
\Omega_o^\prime(r) = \frac{V_{\theta \hskip 0.5pt 0}(r) - V_{\theta \hskip 0.5pt 2}(r)}{r},
\end{equation}
where that $V_{\theta \hskip 0.5pt 0}$ and $V_{\theta \hskip 0.5pt 0}$ are from Equation (5) without the second term on the right-hand side, are consistent with the TW84 method given the size of the 95\% CIs. It is therefore assumed that to a very good approximation
\begin{equation}
V_{\theta \hskip 0.5pt 2}(r) = V_{\theta \hskip 0.5pt 0}(r) - r\,\Omega_o,
\end{equation}
and this is substituted into Equation (5) to break the degeneracy.

To better understand Equation (10), consider that in the FM at the major axis of an elliptical orbit the velocity of the H{\footnotesize I} in the azimuthal direction is $V_{\theta \hskip 0.5pt 0}(r)$ $-$ $V_{\theta \hskip 0.5pt 2}(r)$. Therefore, the instantaneous angular frequency at which the major axis of the elliptical orbit is rotating is [$V_{\theta \hskip 0.5pt 0}(r)$ $-$ $V_{\theta \hskip 0.5pt 2}(r)$]/$r$. In order to avoid the major axis of the elliptical orbit from drifting with respect to the major axis of the oval distortion, these two axes must be rotating at the same angular frequency. In other words, according to the FM,
\begin{equation}
\mbox{cos}(2[\theta - \theta_o])\big|_{\theta = \theta_o} = 1,
\end{equation}
or
\begin{equation}
[\theta(t) - \Omega_o\,t]\big|_{\theta = \theta_o} = 0.
\end{equation}
The functional form of $\theta(t)$ is unknown, but from the FM,
\begin{equation}
\dot{\theta} = \frac{V_{\theta \hskip 0.5pt 0}(r)}{r} - \frac{V_{\theta \hskip 0.5pt 2}(r)}{r}\,\mbox{cos}(\Theta),
\end{equation}
where the dot notation is used to indicate time derivatives. Differentiating the argument in the square brackets of Equation (12) with respect to time gives
\begin{equation}
[\dot{\theta} - \Omega_o]\big|_{\theta = \theta_o} = 0.
\end{equation}
Equation (10) follows from substituting Equation (13) into Equation (14) and evaluating at $\theta$ = $\theta_o$.

It is extremely important to point out that Equation (10) is not derived from the potential, and as such it is not a general result that applies to all bar-like potentials. Furthermore, the FM describes the different velocity components in concentric rings of data and is therefore not a model of the orbit shape. There are many different types of orbit shapes that depend on the properties of a bar-like perturbing potential (e.g., SW93; Hayashi \& Navarro 2006; Binney \& Tremaine 2008, Chapter 3), and thus many possibilities for relating $V_{\theta \hskip 0.5pt 2}$ to different perturbing potentials. It is, however, both unnecessary and beyond the scope of this paper to derive Equation (10) from the potential as long as the dominating components of the H{\footnotesize I} are elliptical orbits and net radial flows, and it can be demonstrated that Equation (9) is consistent with other methods for measuring $\Omega_o$.  Although a derivation from the potential is nontrivial owing to nonconservative forces, the relationship between the FM and the potential is discussed in Section 5.2

After substituting Equation (10) into Equation (5) and rearranging,
\begin{align}
V_y(r,\theta)  - r\,\Omega_{\mbox{\footnotesize o}}\,\mbox{cos}(\Theta)\,\mbox{cos}(\theta) = & \hskip 2pt V_{\theta \hskip 0.5pt 0}(r)\,[1-\mbox{cos}(\Theta)]\,\mbox{cos}(\theta) \nonumber \\ 
& \hskip 2pt + V_{{r} \hskip 0.5pt 0}(r)\,\mbox{sin}(\theta) \nonumber \\ & \hskip 2pt - V_{{r} \hskip 0.5pt 2}(r)\,\mbox{sin}(\Theta)\,\mbox{sin}(\theta).
\end{align}
The matrix of independent variables for the right-hand side of Equation (15) has a rank of 3, so it is possible to find a unique solution for the three remaining velocity variables. Estimates of $V_{\theta \hskip 0.5pt 2}(r)$ are then found from calculating Equation (10) using the results for $V_{\theta \hskip 0.5pt 0}$ and $\Omega_o$. Equation (15) is the modified full model, hereafter referred to as E+R, that is used in  Section 4.4 for finding $V_{{r} \hskip 0.5pt 0}$ in the oval distortion. Table 1 summarizes the adopted and measured variables in E+R. 

\begin{table}[t]
\centering
\caption{Variables in the E+R Model}
\begin{tabular}{lll}
\tablewidth{0pt}
\\[-10.5pt]
\hline
\hline
Variable & Value & Reference \\
\hline
$\alpha_{\mbox{\footnotesize kc}}$  & 12$^{\mbox{\tiny h}}$ 50$^{\mbox{\tiny m}}$ 53\fs0 & (1)\\
$\delta_{\mbox{\footnotesize kc}}$  & +41$^{\circ}$ 07$^{\prime}$ 13\farcs2 & (1)\\
$V_{\mbox{\footnotesize sys}}$  & 306.7 km s$^{-1}$ & (1)\\
$\psi_{\mbox{\footnotesize inc}}$  & 42\fdg0 & (2)\\
$\phi_{\mbox{\footnotesize maj}}$  & 288\fdg7 & (2)\\
$\theta_o$  & 59\fdg8 $\pm$ 1\fdg4 & (2)\\
$\Omega_o$  & 0.67 $\pm$ 0.05 km s$^{-1}$ arcsec$^{-1}$ & (2)\\
\hline
\multicolumn{3}{l}{{\bf References}--(1) db08; (2) this paper, Section 4.}
\end{tabular}
\end{table}

The results for two reduced forms of Equation (5) are provided in Section 4.4 for comparison with the results for E+R. The first reduced model, hereafter referred to as Eonly, describes purely elliptical orbits (e.g., SS07, SS10). It excludes the second term on the right-hand side of Equation (5) that accounts for $V_{{r} \hskip 0.5pt 0}$. The second reduced model, hereafter referred to as C+R, includes $V_{{\theta} \hskip 0.5pt 0}$ and $V_{{r} \hskip 0.5pt 0}$. It excludes the third and fourth terms on the right-hand side of Equation (5) that modify circular orbits into elliptical ones. The results for a model of only $V_{\theta \hskip 0.5pt 0}$ are indistinguishable from the $V_{\theta \hskip 0.5pt 0}$ results for C+R, so they are not shown in this paper. The descriptions of the velocity models are summarized in Table 2.

\subsection{Model for $\theta_o$}

\begin{table}[t]
\centering
\caption{Velocity Models}
\begin{tabular}{ll}
\tablewidth{0pt}
\hline
\hline
Model & Description\\
\hline
FM  & Full model\\
E+R  & Elliptical orbits and net radial flows\\
Eonly & Only elliptical orbits\\
C+R  & Circular orbits and net radial flows\\
\hline
\end{tabular}
\end{table}

There are two steps for measuring $\theta_o$. The first step is to filter the NIR data for 180$^\circ$ rotational symmetry.  The second step is to fit a model that approximates the azimuthal location for the peak intensity of a pattern, $\theta_p$, in the filtered data. The mean of $\theta_p$ in the region of the oval distortion is adopted as $\theta_o$. This paper is primarily interested in $\theta_o$, but radial profiles of $\theta_p$ for the whole disk are useful for discussing the results. 

The NIR data are filtered using the method of Elmegreen et al. (1992).  The method consists of performing the operation 
\begin{equation}
S_{3.6} = I_{3.6} - [I_{3.6} - I_{180}]_{\mbox{\footnotesize T}},
\end{equation}
where $I_{180}$ is $I_{3.6}$ rotated by 180$^\circ$ about the kinematic center. The subscript T indicates that values $<$ 0  for the difference in the square brackets are truncated to 0. The same operation is performed for other tracers of the oval distortion in Appendix B to help facilitate the discussion in Section 5.3.2 about previous estimates of $\theta_o$.

There are three advantages to using data that are filtered in this way. The first is that the results for the filtered data show much less scatter than the results for the unfiltered data, thus increasing the precision of the measurement. The second is that it is less ambiguous than a Fourier analysis, which is sensitive to bias from small asymmetries (e.g., Elmegreen et al. 1993). The third is that it is more thorough in removing the foreground starlight from the Milky Way in comparison to other methods such as Source Extractor (BA96; see panel (a) in Figure 1 for an example). 

Rings of data in $S_{3.6}$ are modeled as
\begin{equation}
S_{3.6}(r,\theta) = \hskip 2pt S_0+S_2(r)\,\mbox{cos}(2[\theta-\theta_p(r)]).
\end{equation}
Equation (17) is transformed from a model that is nonlinear in the unknown variable $\theta_p$, to one that is linear,
\begin{equation}
S_{3.6}(r,\theta) = \hskip 2pt S_0+S_{2x}(r)\,\mbox{cos}(2 \theta)+S_{2y}(r)\,\mbox{sin}(2 \theta),
\end{equation}
using the difference formula for cosine. The value of $\theta_p$ is found from the fitted variables 
\begin{equation}
S_{2x}(r) = S_2(r)\,\mbox{cos}[2\theta_p(r)]
\end{equation}
and
\begin{equation}
S_{2y}(r) = S_2(r)\,\mbox{sin}[2\theta_p(r)]
\end{equation}
by calculating
\begin{equation}
\theta_p(r)=\frac{1}{2}\mbox{tan}^{-1}\frac{S_{2y}(r)}{S_{2x}(r)}.
\end{equation}

This method for determining $\theta_o$ is preferred to other commonly used methods that fit ellipses to isophotes (e.g., Jedrzejewski 1987, hereafter J87) because the statistical significance of the results for ellipse fitting is less straightforward to evaluate. Methods that fit ellipses to isophotes are not maximum likelihood solutions, whereas the methods explained in Appendix A.1 for fitting Equation (18) are. The extra information about ellipticity that is provided by fitting ellipses to isophotes is unnecessary for the purpose of this paper. Methods that fit ellipses to isophotes are discussed in Section 5.3.2.

\subsection{Model for $\Omega_o$}

The general form of the TW84 method is used for finding $\Omega_o$. The TW84 method relates the angular frequency of a pattern, $\Omega_p$, to the observable intensity and velocity of a pattern tracer. The mean of $\Omega_p$ in the region of the oval distortion is adopted as $\Omega_o$. Similar to $\theta_p$, this paper is primarily interested in $\Omega_o$, but radial profiles of $\Omega_p$  for the whole disk are useful for discussing the results. 

The original TW84 method assumes that the disk is flat, the tracer obeys mass conservation in the continuity equation, and $\Omega_p$ is a constant function of radius. The relationship derived by TW84 is
\begin{equation}
\mathcal{V} = \Omega_p\, \mathcal{X},
\end{equation}
where for the H{\footnotesize I} data
\begin{equation}
\mathcal{V}_i = \int^{\infty}_{-\infty}I_{HI}(x,y_i)\,V_y(x,y_i)\, dx
\end{equation}
and
\begin{equation}
\mathcal{X}_i = \int^{\infty}_{-\infty}I_{HI}(x,y_i)\,x\, dx.
\end{equation}
Equation (22) is derived by integrating the continuity equation over an area of the disk bounded by $-\infty$ $<$ $x$ $<$ $\infty$ and $y_i$ $<$ $y$ $<$ $\infty$, or similarly for negative $y$. For multiple calculations of Equations (23) and (24) at different $y_i$, $\Omega_p$ is the slope of a line fit through a plot of $\mathcal{V}(\mathcal{X})$ that has an intercept of $\mathcal{V}(0)$ = 0.

It is assumed that the optical disk ($r$ $\leqslant$ $R_{\mbox{\footnotesize 25}}$) of NGC 4736 is approximately flat. There are $\approx$10$^\circ$ changes in both $\psi_{\mbox{\footnotesize inc}}$ and $\phi_{\mbox{\footnotesize maj}}$ beyond $R_{\mbox{\footnotesize 25}}$ for titled rings (dB08; MVD93), but it is unclear whether this is due to a warp, the uniqueness problem for nonlinear least squares, excluding noncircular velocity components in the fitted model, or some combination of these reasons. The effect of a warp is minimized by restricting the application of the method to $|y_i|$ $\leqslant$ $R_{\mbox{\footnotesize 25}}$. Any effect a warp may have on the results is tested for in Section 4.3.

A tracer of the pattern that is also a tracer of a component of the interstellar medium (ISM), such as the H{\footnotesize I}, will not satisfy the assumption of mass conservation in the strictest sense. This assumption may be unnecessary for ISM tracers in secularly evolving galaxy disks. When the source function is included in the continuity equation, it can be shown that it is reasonable to assume that sources and sinks for ISM tracers have a negligible effect on the measured value of $\Omega_p$ (Westpfahl 1998; Speights \& Westpfahl 2011, 2012). The effect of violating the assumption of mass conservation will manifest as nonzero intercepts, or deviations from linearity, in plots of $\mathcal{V}(\mathcal{X})$, and these deviations are often statistically insignificant (see SR16 for a discussion). Examples of linear trends with approximately zero intercepts in $\mathcal{V}(\mathcal{X})$ for ISM tracers are shown in Rand \& Wallin (2004, hereafter RW04), Hernandez et al. (2005), Emsellem et al. (2006), Fathi et al. (2007, 2009), Chemin \& Hernandez (2009), Gabbasov et al. (2009), Banerjee et al. (2013), and SR16. 

The approximately zero intercepts are explainable by noting that the TW84 method integrates the continuity equation over an area of the disk. This area will include a combination of sources and sinks. Approximately zero intercepts imply that the sum of the sources and sinks is much smaller than the left- and right-hand sides of Equation (22).

The assumption that $\Omega_p$ is a constant function of radius across the whole disk is the most uncertain of the three assumptions. This is well established from direct measurements (Westpfahl 1998; Hernandez et al. 2005; Merrifield et al 2006, hereafter M06; Meidt et al. 2008a, 2008b, 2009; Fathi et al. 2009; Speights \& Westpfahl 2011; Speights \& Westpfahl 2012; SS16), simulations (e.g., Sellwood \& Sparke 1988, Sellwood \& Carlberg 2014), and observations (e.g., SSW93, and references therein).  The uncertainty in assuming that $\Omega_p$ is constant across the whole disk is discussed in Binney \& Tremaine (2008, Chapter 6), and Dobbs \& Baba (2014, and references therein). In Sections 2 and 4.1 it is shown that NGC 4736 contains multiple patterns of different types. Each of these may have different values of $\Omega_p$. 

The general form of the TW84 method allows for radial variation in $\Omega_p$.
When $\Omega_p$ is allowed to vary with radius, the result from integrating the continuity equation is a Volterra equation of the first kind,
\begin{eqnarray}
\mathcal{V}_i = \int^{\infty}_{y_i}\Omega_p(r)\{I_{HI}(\sqrt{r^2-y_i^2},y_i) - & \nonumber \\ & \hskip -70pt I_{HI}(-\sqrt{r^2-y_i^2},y_i)\}\,r\,dr
\end{eqnarray}
(Engstr\"oem 1994, M06). Fits of Equation (25) are unstable to noise (E94, M06, Meidt 2008a). Stable fits are found using Tikhonov regularization (Aster et al. 2012, Chapter 5, hereafter A12). The procedure for using Tikhonov regularization, and the associated L-curve criteria for determining the strength of the regularization $\lambda$, are explained in  Appendix A.1.

\section{Results} \label{sec:res}

The mathematical models are fit to the data using standard least-squares methods.  A brief explanation of the different methods is provided in  Appendix A.1.  The uncertainties calculated in this paper are 95\% CIs. The uncertainties are explained in Appendix A.2. 

\subsection{Results for $\psi_{\mbox{\footnotesize inc}}$ and $\phi_{\mbox{\footnotesize maj}}$}

\begin{figure}[ht!]
\centering
\includegraphics[width=0.97\columnwidth]{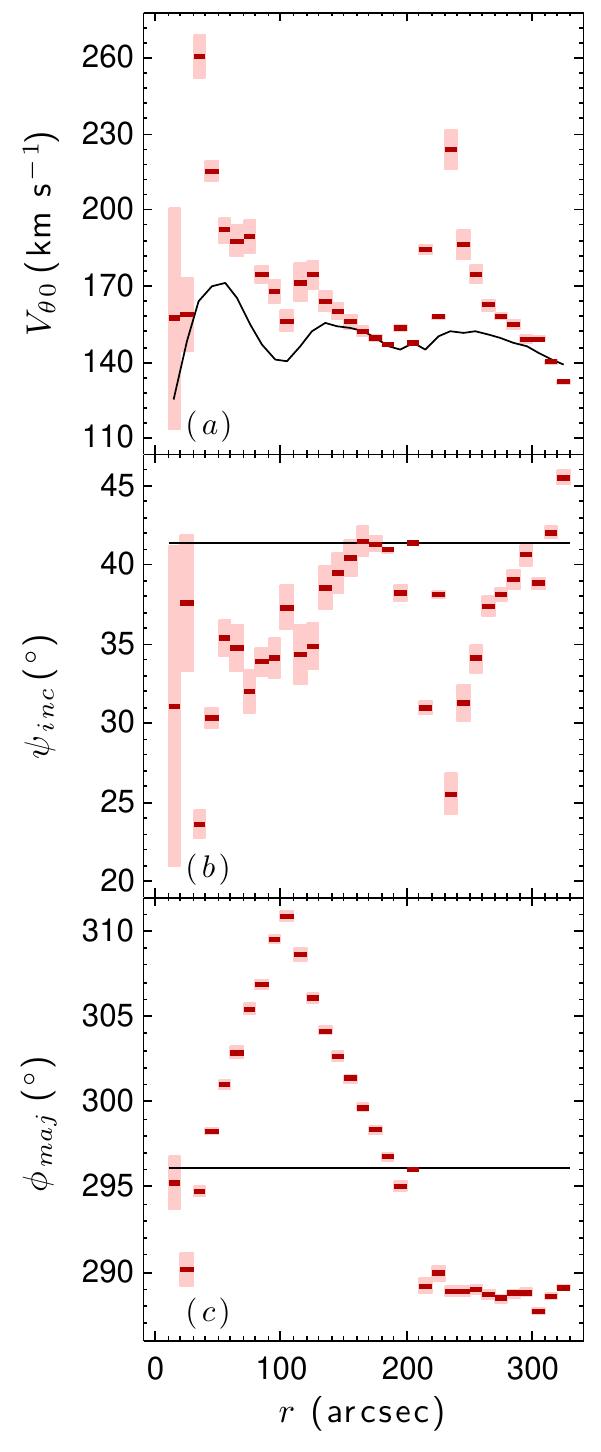} 
\caption{Results for when $\psi_{\mbox{\footnotesize inc}}$ and $\phi_{\mbox{\footnotesize maj}}$ are allowed to vary with radius. The dark-red solid line segments and light-red shading show the results for each ring of data and the 95\% CIs, respectively. The black solid lines show the starting values for the Levenberg-Marquardt algorithm.}
\end{figure}

The values for $\psi_{\mbox{\footnotesize inc}}$ and $\phi_{\mbox{\footnotesize maj}}$ are found from the results of fitting Equation (4) to the region between the end of the oval distortion and $R_{25}$. This region lacks a known engine for secular evolution, and shows an approximately constant trend in the radial profile of $\phi_{\mbox{\footnotesize maj}}$ when it is allowed to vary with radius. 

The results for when $\psi_{\mbox{\footnotesize inc}}$ and $\phi_{\mbox{\footnotesize maj}}$ are allowed to vary with radius are demonstrated using the Levenberg-Marquardt algorithm applied to Equation (4) with a weighting function of $|$cos($\theta$)$|$. Fits are found in 10$^ {\prime\prime}$ rings for the region 10$^{\prime\prime}$ $\leqslant$ $r$  $\leqslant$ 330$^{\prime\prime}$. The starting values of $\psi_{\mbox{\footnotesize inc}}$ = 41\fdg4 and $\phi_{\mbox{\footnotesize maj}}$ = 296\fdg1 are adopted from the mean values found by dB08. The starting values for $V_{\theta \hskip 0.5pt 0}$ are from the results of using the normal equations for Equation (4), with $\psi_{\mbox{\footnotesize inc}}$ and $\phi_{\mbox{\footnotesize maj}}$ set to the mean values found by dB08. The starting value of the dampening factor in the algorithm, $\Lambda$, is 1 $\times$ 10$^{-2}$, and is increased or decreased by a factor of 10 (see Appendix A.1). The algorithm is allowed to continue until it converges to a tolerance of 1 $\times$ 10$^{-5}$.

Figure 2 shows the results. They are in excellent agreement with those shown by dB08 in their Figure 80. The results for $V_{\theta \hskip 0.5pt 0}$ and $\psi_{\mbox{\footnotesize inc}}$ show the uniqueness problem for this nonlinear model. The peaks in the radial profile of $V_{\theta \hskip 0.5pt 0}$ correspond to dips in $\psi_{\mbox{\footnotesize inc}}$. An increase or decrease in $\psi_{\mbox{\footnotesize inc}}$ is compensated for by an opposite change in $V_{\theta \hskip 0.5pt 0}$. The radial profile for the starting value of $V_{\theta \hskip 0.5pt 0}$ when $\psi_{\mbox{\footnotesize inc}}$ is held constant  shows much less scatter than the fitted results. The radial profile of $\phi_{\mbox{\footnotesize maj}}$ is relatively flat in the region between the end of the oval distortion and R$_{25}$. The mean values in this region are $\psi_{\mbox{\footnotesize inc}}$ = 39\fdg1 $\pm$ 3\fdg6 and $\phi_{\mbox{\footnotesize maj}}$ = 288\fdg7 $\pm$ 0\fdg4.

These mean values for $\psi_{\mbox{\footnotesize inc}}$ and $\phi_{\mbox{\footnotesize maj}}$ may not represent a global minimum in the SSRs owing to the uniqueness problem. They are refined by calculating the SSRs from fits of Equation (4) to the region 220$^ {\prime\prime}$ $\leqslant$ $r$  $\leqslant$ 330$^ {\prime\prime}$ that assume different combinations of constant values for $\psi_{\mbox{\footnotesize inc}}$ and $\phi_{\mbox{\footnotesize maj}}$. The fits are performed using the normal equations. The only variable fitted for, $V_{\theta \hskip 0.5pt 0}$, is allowed to vary from ring to ring in 10$^{\prime\prime}$ rings. A $|$cos($\theta$)$|$ weighting function is used for giving greater weight to the kinematic major axis. 

\begin{figure}[t!]
\centering
\includegraphics[width=1\columnwidth]{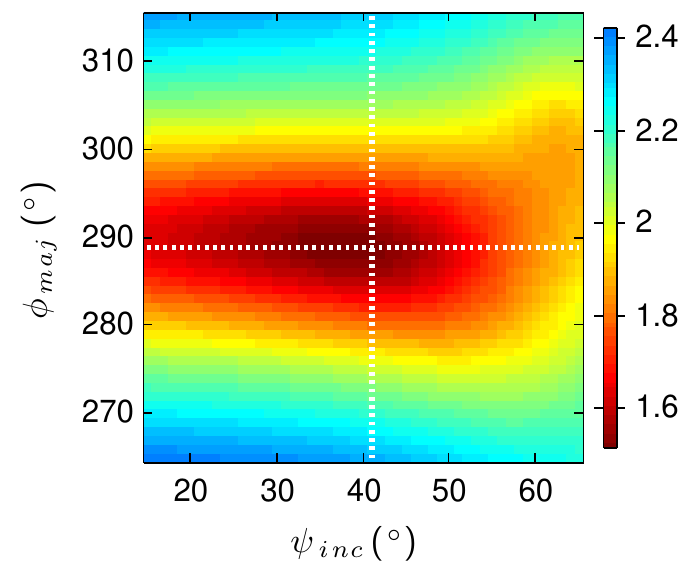} 
\caption{Map of the SSRs for different combinations of $\psi_{\mbox{\footnotesize inc}}$ and $\phi_{\mbox{\footnotesize maj}}$ when the normal equations are used for finding $V_{\theta \hskip 0.5pt 0}$ in Equation (4). The results are log scaled. The color bar is in units of log(km$^2$ s$^{-4}$). The vertical and horizontal white dotted lines show the adopted values of $\psi_{\mbox{\footnotesize inc}}$ and $\phi_{\mbox{\footnotesize maj}}$, respectively, that intersect at the global minimum in the SSRs.}
\end{figure}

The SSR results for 15$^{\circ}$ $\leqslant$ $\psi_{\mbox{\footnotesize inc}}$ $\leqslant$ 65$^{\circ}$ and 265$^{\circ}$ $\leqslant$ $\phi_{\mbox{\footnotesize maj}}$ $\leqslant$ 315$^{\circ}$ are shown in Figure 3. There is a clearly defined global minimum in the SSRs at $\psi_{\mbox{\footnotesize inc}}$ = 42\fdg0 and $\phi_{\mbox{\footnotesize maj}}$ = 288\fdg7. These values are adopted for the rest of this paper. The adopted value of $\phi_{\mbox{\footnotesize maj}}$ is the same as the mean value for tilted rings between the end of the oval distortion and R$_{25}$. The adopted value of $\psi_{\mbox{\footnotesize inc}}$ is consistent with the mean value for titled rings in the same region given the size of the 95\% CI of the mean value for titled rings.

Uncertainties are not reported for the adopted values of $\psi_{\mbox{\footnotesize inc}}$ and $\phi_{\mbox{\footnotesize maj}}$. The method for estimating $\psi_{\mbox{\footnotesize inc}}$ and $\phi_{\mbox{\footnotesize maj}}$ assumes values for them instead of fitting for them, and this invalidates the methods in this paper for estimating uncertainties. The uncertainties in the literature for the type of high-quality data in this paper are typically a few degrees or less, as demonstrated by the 95\% CIs for the means of the results for tilted rings between the end of the oval distortion and R$_{25}$. The accuracy of the adopted values for $\psi_{\mbox{\footnotesize inc}}$ and $\phi_{\mbox{\footnotesize maj}}$ is discussed in  Section 5.3.1.

\begin{figure}[t!]
\centering
\includegraphics[width=1\columnwidth]{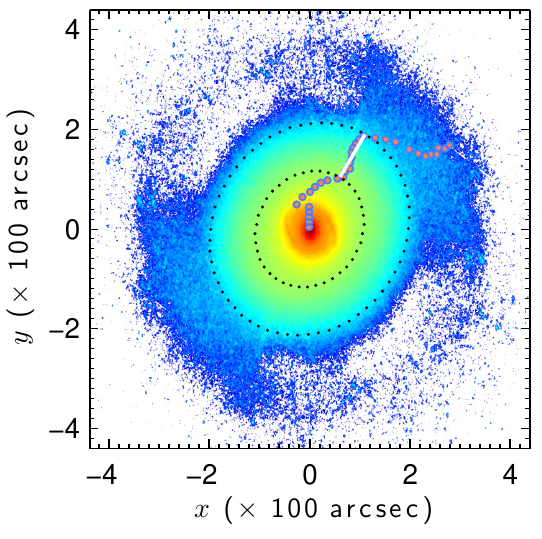} 
\caption{Map of $S_{3.6}$ rotated and flipped upside down to align the galaxy $x$- and $y$-axes horizontally and vertically, respectively. The map is log scaled. The values less than 0 are background noise and are set to 0 in the figure. Removing the background noise and the chosen color scheme help distinguish the different patterns. The black dotted lines trace the inner and outer radius of the oval distortion. The red dots that are outlined in blue show the results for $\theta_p$. The white line shows the mean of the $\theta_p$results in the region of the oval distortion.}
\end{figure}
%

\begin{figure*}[ht!]
\centering
\includegraphics[width=1\textwidth]{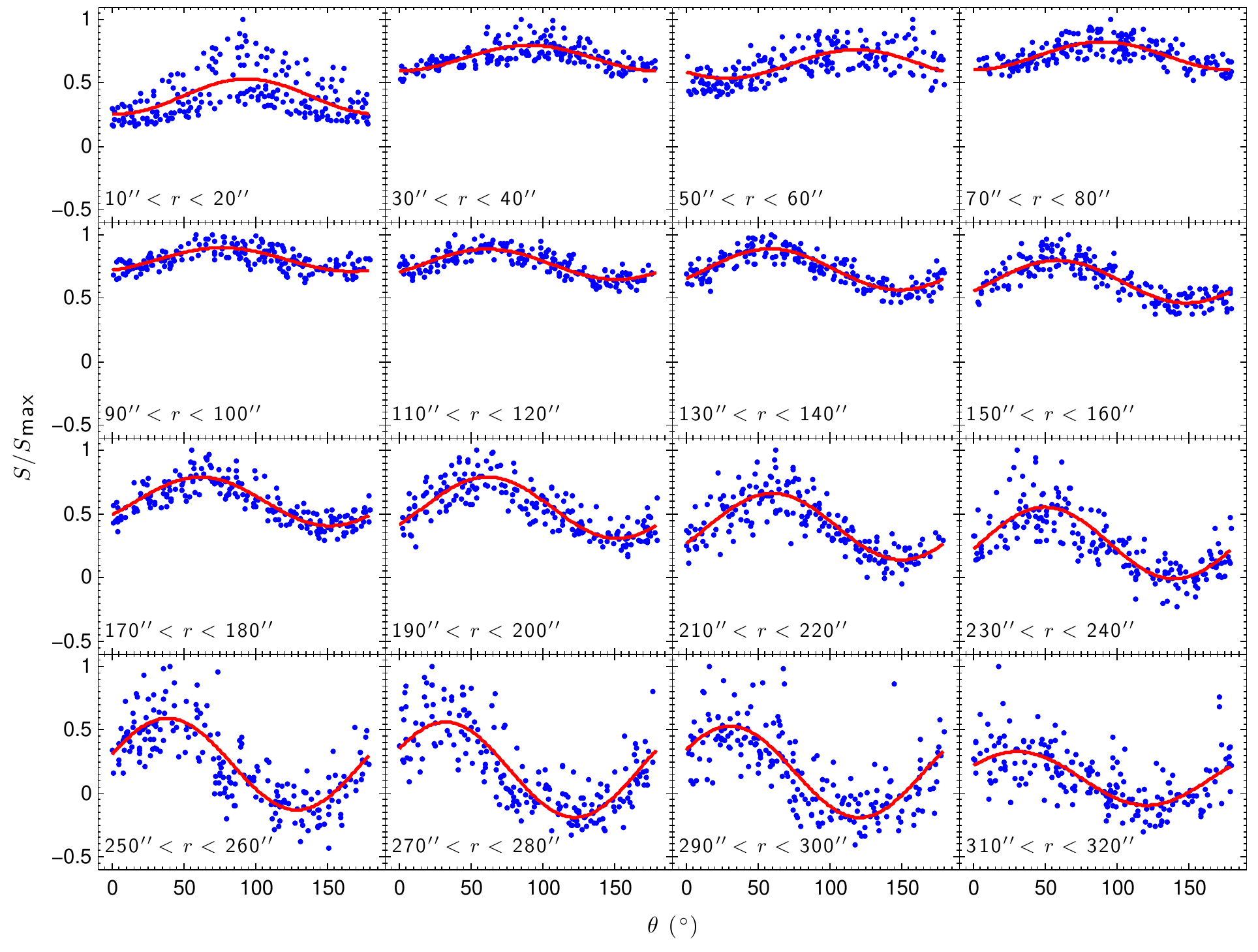} 
\caption{Results for fits of Equation (14) that are used for calculating $\theta_p$ in Equation (17). Only data and results for 0$^{\circ}$ $\leqslant$ $\theta$ $<$ 180$^\circ$ are shown owing to the 180$^{\circ}$ symmetry of $S_{3.6}$. The blue points show $S_{3.6}$ normalized by the maximum value in a ring of data, $S_{\mbox{\footnotesize max}}$. Only 200 randomly chosen points are shown out of the total number of points for clarity. The red solid lines show the fitted models. The range in $r$ for the ring annuli of the fits are shown in the bottom left of each panel. Every other ring of data is shown for brevity.}
\end{figure*}

\subsection{Results for $\theta_o$}

The value of $\theta_o$ is found from the results for fits of Equation (18) that are used in calculations of Equation (21). The fits are performed for 10$^{\prime\prime}$ rings in the region 0$^{\prime\prime}$ $\leqslant$ $r$ $\leqslant$ 330$^{\prime\prime}$. This range in $r$ is chosen to check how well the mathematical form of Equation (18) distinguishes the different patterns in NGC 4736. The fits are restricted to 0$^{\circ}$ $\leqslant$ $\theta$ $<$ 180$^\circ$ to account for the data in $S_{3.6}$ repeating after 180$^{\circ}$.

Figure 4 shows $S_{3.6}$ in galaxy coordinates. The many patterns of NGC 4736 are distinguishable in the figure. The nuclear bar is aligned closely to the $y$-axis. The major axis of the oval distortion is tilted between the +$x$- and +$y$-axes. Between the nuclear bar and the oval distortion, the spiral patterns making up the inner pseudoring are clearly visible. The less complete outer pseudoring is explainable by noting that its spiral patterns are more open and less symmetric than the ones making up the inner pseudoring. The asymmetry of the outer pseudoring's spiral patterns is observable in the $I_{HI}$ data in panel ($b$) of Figure 1 (see also T09 for the outer spiral patterns in other wavelengths).

Figure 5 provides examples of the results for fits of Equation (18). The results follow the peaks and troughs in the azimuthal profiles of $S_{3.6}$, especially in the region of the oval distortion. The azimuthal profiles of $S_{3.6}$ in the region of the oval distortion only show one peak, consistent with the assumption of a single oval distortion in this region. The peaks are sharper in the nuclear region, $r$ $\leqslant$ 40$^{\prime\prime}$. There is more scatter in $S_{3.6}$ for the region beyond the end of the oval distortion.

Figure 6 shows the radial profile of $\theta_p$ from calculations of Equation (21). The mean of $\theta_p$ in the region of the oval distortion is 59\fdg8 $\pm$ 1\fdg4. Most of the CIs for $\theta_p$ fit within the CI for the mean in the region of the oval distortion, consistent with the assumption of a single oval distortion in this region. This mean is adopted as the estimate for $\theta_o$ in E+R and Eonly. The mean of $\theta_p$ in the nuclear region is 91\fdg7 $\pm$ 8\fdg3. This is adopted as the estimate of the phase angle for the location of the nuclear bar, $\theta_b$, which is useful for interpreting the results for $\Omega_p$ in Section 4.3, and discussing the accuracy of the results for $\theta_o$ in Section 5.3.2.

\begin{figure}[ht!]
\includegraphics[width=1\columnwidth]{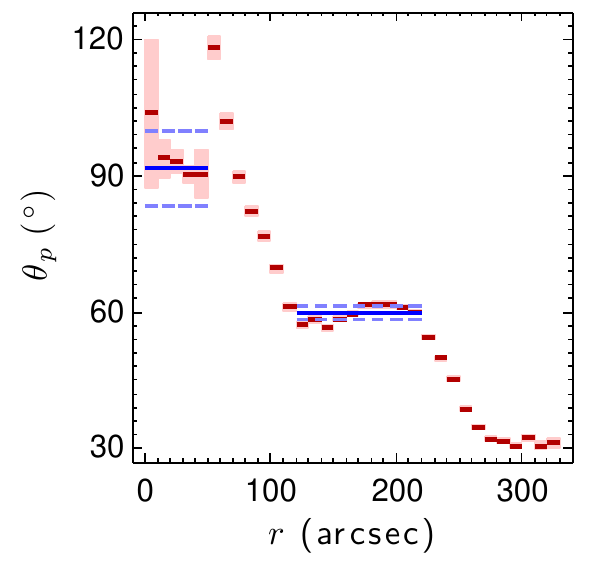} 
\caption{Results for calculations of Equation (17). The dark-red solid line segments and light-red shading show the results for $\theta_p$ and the 95\% CIs, respectively. The blue solid lines and blue dashed lines show the mean of $\theta_p$ and the 95\% CIs, respectively, for the nuclear bar and oval distortion.}
\end{figure}

\subsection{Results for $\Omega_o$}

The value of $\Omega_o$ is found from the results for fits of Equation (25). To prepare the H{\footnotesize I} data for the fits, they are first rotated to align the pixel gridding parallel to the $x$- and $y$-axes. Next, the integrands are calculated from the rotated data. Integration is then performed by summing pixels in the integrands along paths that are parallel to the $x$-axis. 

The fits are performed for 10$^{\prime\prime}$ rings, except for the outermost ring of data. The outermost ring extends from 330$^{\prime\prime}$ to the edge of the data so that the integrals converge as they do in the derivation by TW84. The effect of any warping is found to be negligible by comparing the results with those that exclude the outermost ring. This affects the results for a few of the remaining outer rings, but the effect is smaller than the 95\% CIs. The results for the outermost ring are excluded from the figures for $\Omega_p$. They are unnecessary for the purpose of this paper, and including them unnecessarily increases the $r$-axis. Their values are less than the results for the next adjacent ring inward, consistent with the generally decreasing trends that are found. 
 
Figure 7 shows the results for a fit of Equation (25) across the whole disk and the L-curve for determining $\lambda$. Possible locations for corotation and Lindblad resonance are included with $\Omega_p$ in panel (b). The value of $\Omega$ = $V_{\theta \hskip 0.5pt 0}$/$r$ is calculated from the results for C+R. There is bias in $V_{\theta \hskip 0.5pt 0}$ when $V_{\theta \hskip 0.5pt 2}$ and $V_{r \hskip 0.5pt 2}$ are excluded from the fitted model (Section 4.4), but the effect is small when dividing by $r$ to calculate $\Omega$, especially near corotation. 

\begin{figure}[t]
\includegraphics[width=1\columnwidth]{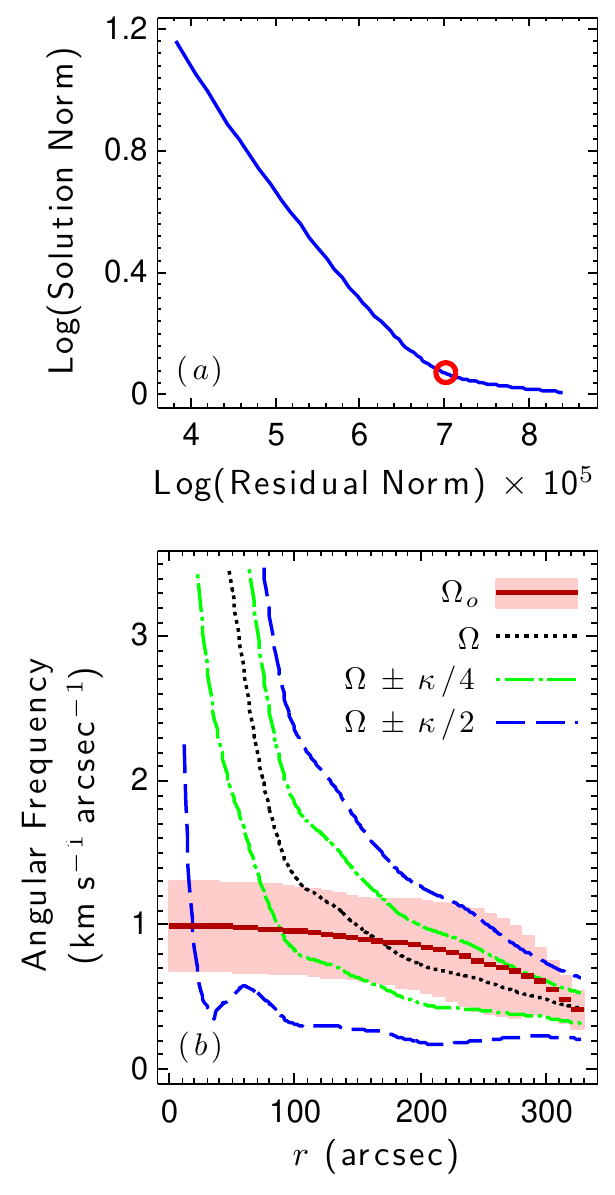} 
\caption{Results for the general form of the TW84 method. Panel ($a$) shows a solid blue line for the L-curve. The center of the red circle in panel ($a$) indicates the corner of the L-curve for determining $\lambda$. Panel ($b$) shows dark-red solid line segments and light-red shading for $\Omega_p$ and the 95\% CIs, respectively. Included in panel ($b$) are possible locations for corotation and Lindblad resonances. The lines for the possible locations of resonance are defined in the legend in panel ($b$).}
\end{figure}

The value of $\Omega_o$ is poorly constrained by the results in panel ($b$) of Figure 7. An approximately constant value of $\Omega_p$ for most of the inner part of the optical disk is consistent with the results given the size of the 95\% CIs. The 95\% CIs are too large, however, to rule out alternative functional forms for $\Omega_p$ such as a linear one, a quadratic one, or some combination of these. If the 95\% CIs are ignored, there is a trend in $\Omega_p$ that decreases with increasing radius. Furthermore, using the mean value of $\Omega_p$ in Figure 7 for the region of the oval distortion as an estimate of $\Omega_o$ produces values of $V_{{r} \hskip 0.5pt 0}$ that are $\approx$ -30 km s$^{-1}$, which are much larger than what is expected (e.g., paragraph 2 of the Introduction).

The results in panel ($b$) of Figure 7 are explainable if the nuclear bar is rotating more quickly than the oval distortion, and by noting that regularization penalizes discontinuities in the functional form of $\Omega_p$. A faster-rotating nuclear bar is allowed for in the $\theta_o$ and $\theta_b$ results. The 31\fdg9 $\pm$ 8\fdg4 difference between $\theta_o$ and $\theta_b$ is inconsistent with the nuclear bar belonging to $x_2$-type orbits rotating at the same angular frequency as the $x_1$-type orbits of the oval distortion (C80).

If there are two distinct values of $\Omega_p$ for the nuclear bar and oval distortion, then the penalties imposed by regularization will bias the results. In the presence of a faster rotating nuclear bar, the smaller values of $\Omega_p$ in the outer region of the galaxy will bias the results in the inner region toward 0 km s$^{-1}$ arcsec$^{-1}$. Likewise, the larger values of $\Omega_p$ in the inner region of the galaxy will bias the results in the outer region toward infinity. 

The L-curve criteria are meant to minimize this type of bias, but differences in the signal-to-noise ratio in the inner and outer parts of the disk complicate its effectiveness.  As the intensity decreases with increasing radius, so does the signal-to-noise ratio. The amount of regularization needed for stabilizing the solution therefore increases with increasing radius.  

This explanation is demonstrated in Figure 8 using a value of $\lambda$ that is 10\% of the one for the results shown in Figure 7.  The results in Figure 8 show an approximately constant $\Omega_p$ for the nuclear bar and most of the inner pseudoring and oscillations from solution instabilities that propagate outward from the inner radius of the oval distortion. The mean of the results for $r$ $\leqslant$ 40$^{\prime\prime}$ is 2.96 $\pm$ 0.36 km s$^{-1}$ arcsec$^{-1}$. This is adopted as the estimate for the angular frequency of the nuclear bar, $\Omega_b$. Although the 95\% CIs are large in the nuclear region, this estimate for $\Omega_b$ is useful for discussing the accuracy of the results for $\Omega_o$ in Section 5.3.3.

\begin{figure}[t!]
\includegraphics[width=1\columnwidth]{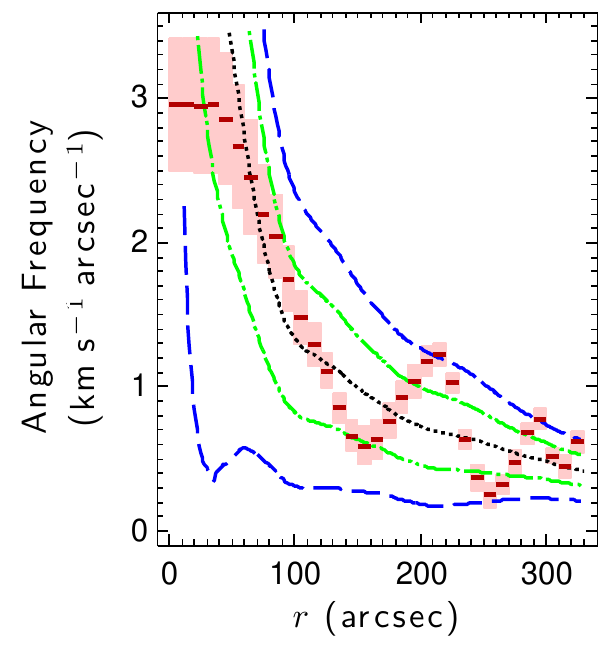} 
\caption{Results for the general form of the TW84 method with less regularization. The figure is formatted in the same way as panel ($b$) in Figure 7.}
\end{figure}

The bias from the larger values of $\Omega_p$ in the region $r$ $<$ 120$^{\prime\prime}$ is removed by excluding integration paths that pass over that region (e.g., SR16). A fit for 120$^{\prime\prime}$ $\leqslant$ $|y_i|$ $\leqslant$ R$_{\mbox{\footnotesize 25}}$ is shown in Figure 9. The results in Figure 9 for the region of the oval distortion are much better described by an approximately constant value for $\Omega_o$ than the results in Figure 7. The oval distortion begins near an inner ultraharmonic (4:1) Lindblad resonance (green dashed-dotted line in Figure 9), and extends up to a corotation resonance (black dotted line in Figure 9).  The locations of resonances are consistent with the assumption of $x_1$-type orbits for the oval distortion. The mean of $\Omega_p$ for the oval distortion is 0.67 $\pm$ 0.05 km s$^{-1}$ arcsec$^{-1}$. This is adopted as the estimate for $\Omega_o$ on the left-hand side of the E+R model. 

\begin{figure}[t!]
\includegraphics[width=1\columnwidth]{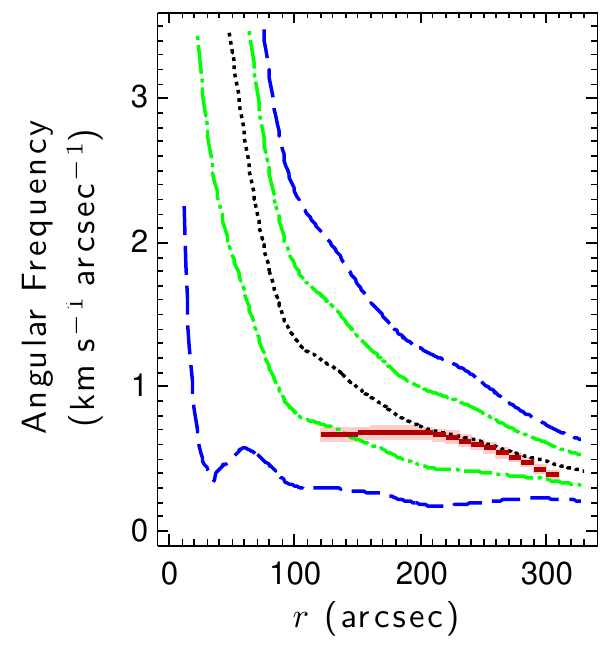} 
\caption{Results for the general form of the TW84 method that excludes integrals in the region $|y_i|$ $<$ 120$^{\prime\prime}$. The figure is formatted in the same way as panel ($b$) in Figure 7. }
\end{figure}
%
\begin{figure*}[t!]
\includegraphics[width=1\textwidth]{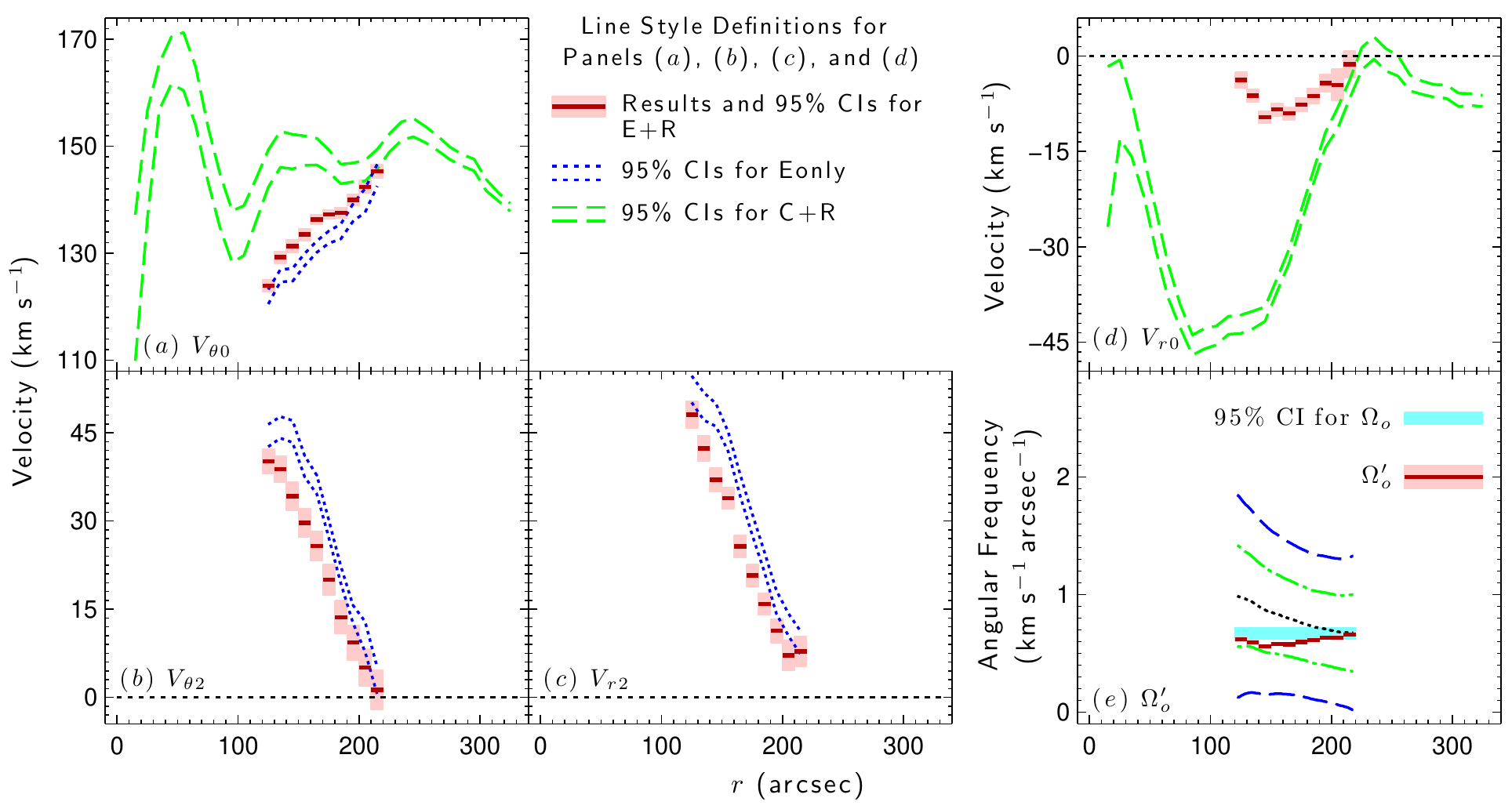} 
\caption{Results for modified and reduced velocity models. The fitted or calculated variables are indicated in the bottom left of each panel. The dark-red solid line segments and light-red shading in panels ($a$)--($d$) show the results for E+R and the 95\% CIs, respectively. The dotted blue lines in panels ($a$)--($c$) show the 95\% CIs for Eonly. The green dashed lines in panels ($a$) and ($d$) show the 95\% CIs for C+R. These line style definitions for panels ($a$) -- ($d$) are summarized in the top center of the figure for convenience. The black dotted lines at 0 km s$^{-1}$ in panels ($b$)--($d$) are provided for reference. The dark-red line segments in panel ($e$) show $\Omega_o^\prime$ calculated from the results for Eonly. The possible locations for resonance in this panel are formatted in the same way as panel ($b$) in Figure 7, but are calculated using $V_{{\theta} \hskip 0.5pt o}$ from Eonly. The cyan shading in panel ($e$) shows the 95\% CI for $\Omega_o$ for comparison with $\Omega_o^\prime$.}
\end{figure*}

\begin{figure}[t!]
\includegraphics[width=1\columnwidth]{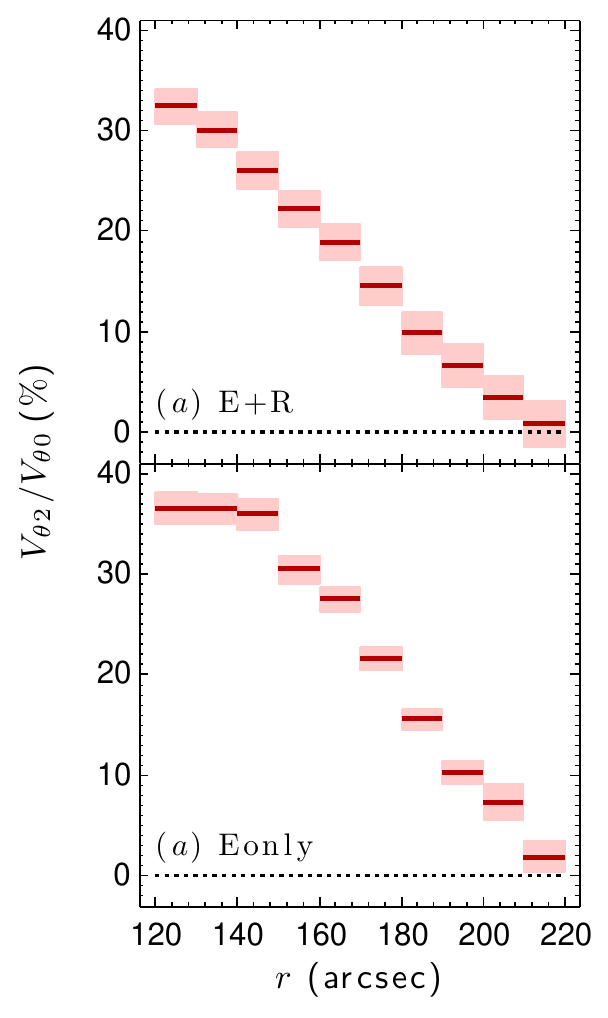} 
\caption{Percent ratio of  $V_{{\theta} \hskip 0.5pt 2}$ to $V_{{\theta} \hskip 0.5pt 0}$. Panels ($a$) and ($b$) show the results for E+R and Eonly, respectively. The dark-red line segments and light-red shading show the ratio of $V_{{r} \hskip 0.5pt 2}$ to $V_{{r} \hskip 0.5pt 0}$ and the 95\% CIs, respectively. The black dotted lines at 0\% are provided for reference.}
\end{figure}

\subsection{Results for the Velocities}

Figure 10 shows the results for the three velocity models summarized in Table 2. All three models are fit in 10$^{\prime\prime}$ rings. The E+R and Eonly models are fit to the region 120$^{\prime\prime}$ $\leqslant$ $r$ $\leqslant$ 220$^{\prime\prime}$. The C+R model is fit to the region 10$^{\prime\prime}$ $\leqslant$ $r$ $\leqslant$ 330$^{\prime\prime}$. 

Excluding velocity components in the reduced models will bias the results for those models. Included in this subsection are estimates of the mean bias. This is calculated by finding the mean values of the results for E+R and then subtracting them from the mean values of the results for C+R and Eonly. For C+R, the calculation of the mean values only include the results in the region of the oval distortion. For example, to find the mean bias in the $V_{\theta \hskip 0.5pt 0}$ results for C+R, the mean of the $V_{\theta \hskip 0.5pt 0}$ results for E+R is subtracted from the mean of the $V_{\theta \hskip 0.5pt 0}$ results for C+R in the region of the oval distortion. Positive bias corresponds to results for a reduced model that are larger than the results for E+R, and likewise for negative bias.

Panel ($a$) shows the results for $V_{\theta \hskip 0.5pt 0}$. The results for E+R and Eonly in the region of the oval distortion remove most of the bump in the shape of the radial profile for C+R there. The mean bias in the $V_{\theta \hskip 0.5pt 0}$ results for C+R in the region of the oval distortion is 11.5 $\pm$ 5.1 km s$^{-1}$. The mean bias in the $V_{\theta \hskip 0.5pt 0}$ results for Eonly is -2.4 $\pm$ 0.9 km s$^{-1}$. The results for E+R and Eonly are more similar to each other at the beginning and ending of the oval distortion given the size of the 95\% CIs.

Panels ($b$) and ($c$) show the results for $V_{\theta \hskip 0.5pt 2}$ and $V_{{r} \hskip 0.5pt 2}$, respectively. The mean bias in the $V_{\theta \hskip 0.5pt 2}$ results for Eonly is 6.7 $\pm$ 2.0 km s$^{-1}$. The mean bias in the $V_{r \hskip 0.5pt 2}$ results for Eonly is 6.4 $\pm$ 2.1 km s$^{-1}$. Similar to the results shown in panel ($a$), both sets of results for E+R and Eonly are more similar to each other at the beginning and ending of the oval distortion given the size of the 95\% CIs. Both sets of results for E+R and  Eonly trend toward 0 km s$^{-1}$ near the end of the oval distortion, as expected for $x_1$-type orbits. 

Panel ($d$) shows the results for $V_{r \hskip 0.5pt 0}$. The mean bias in the $V_{r \hskip 0.5pt 0}$ results for C+R in the region of the oval distortion is -20.3 $\pm$ 8.9 km s$^{-1}$. Both results for E+R and C+R trend toward 0 km s$^{-1}$ near the end of the oval distortion, consistent with the results in panels ($a$)--($c$). The results for C+R show inward net radial flow velocities as large as -45 km s$^{-1}$, which is much larger than what is expected. 

The mean of $V_{{r} \hskip 0.5pt 0}$ for E+R is $\overline{V}_{{r} \hskip 0.5pt 0}$ = -6.1 $\pm$ 1.9 km s$^{-1}$. This is adopted in the rest of this paper as the measured value of $\overline{V}_{{r} \hskip 0.5pt 0}$. The robustness of this result is checked by applying the same procedure separately to receding and approaching halves of the galaxy. The mean of $V_{{r} \hskip 0.5pt 0}$ is $\overline{V}_{{r} \hskip 0.5pt 0}$ = -6.4 $\pm$ 1.8 km s$^{-1}$ and -5.8 $\pm$ 2.8 km s$^{-1}$ for the receding and approaching halves, respectively. These are both consistent with the adopted value given the size of the 95\% CIs.

Panel (e) shows Equation (9) calculated from the results for Eonly for comparison with $\Omega_o$ found using the general TW84 method. Note that in Equation (10) $\Omega_o$ is the mean of $\Omega_p$ from the TW84 method, which is expected to be an approximately constant function of radius for the oval distortion. The calculation of $\Omega^\prime_o$ in Equation (9), however, is a rough estimate of $\Omega_o$ at different radii assuming that net radial flows are negligible.  The radial profile of the $\Omega^\prime_o$ results shows more variation than that of the $\Omega_p$ results from the TW84 method because of the bias in Eonly.

The mean of $\Omega_o^\prime$ is 0.60 $\pm$ 0.04 km s$^{-1}$ arcsec$^{-1}$, consistent with the adopted value of $\Omega_o$ = 0.67 $\pm$ 0.05 km s$^{-1}$ given the size of the 95\% CIs. The values of $\Omega_o^\prime$ are the most similar to $\Omega_o$ near the beginning and ending of the oval distortion, consistent with the results in panels ($a$)--($c$). The largest differences between $\Omega_o$ and $\Omega_o^\prime$ occur in the middle of the oval distortion, coinciding with  the largest values of $V_{r \hskip 0.5pt 0}$ for E+R in panel ($d$). The similarity of the results for $\Omega_o^\prime$ to those for $\Omega_o$ justifies the use of Equation (8) for eliminating $V_{\theta \hskip 0.5pt 2}$ in Equation (5).

The $V_{\theta \hskip 0.5pt 0}$, $V_{\theta \hskip 0.5pt 2}$, and $V_{r \hskip 0.5pt 2}$ results for E+R and Eonly that are shown in panels ($a$)--($c$) also help justify the use of Equation (10) for eliminating $V_{\theta \hskip 0.5pt 2}$ in Equation (5). Both sides of the equal signs for E+R and Eonly are quite different, yet the radial profiles of their results for these three velocities are very similar, as they should be for the small net radial flow velocities in E+R. Most significantly, the radial profiles of $V_{\theta \hskip 0.5pt 2}$ are similar to those of $V_{r \hskip 0.5pt 2}$. This is despite $V_{r \hskip 0.5pt 2}$ being fitted for in both models, whereas $V_{\theta \hskip 0.5pt 2}$ is fitted for in Eonly, and $V_{\theta \hskip 0.5pt 2}$ is calculated from Equation (10) in E+R.

\section{Discussion} 

\subsection{Interpretation of the Results for $V_{{r} \hskip 0.5pt 0}$}

The interpretation of the results for $V_{{r} \hskip 0.5pt 0}$ requires several considerations.   The FM assumes that the dominating velocity components of the H{\footnotesize I} are elliptical orbits and net radial flows, with the major axis of the elliptical orbits aligned with the major axis of the oval distortion. It does not take into account more complicated flow geometries that would result from more complicated potentials. The method relies on accurate estimates of other variables such as $\psi_{\mbox{\footnotesize inc}}$, $\phi_{\mbox{\footnotesize maj}}$, $\theta_o$, and $\Omega_o$. Standard hypothesis testing techniques for comparing the goodness of fit between E+R and the reduced models are inappropriate because of the degeneracy in the FM and the difference in the column matrix of the data, ${\boldsymbol d}$, on the left-hand side of Equations (5) and (15). This is why that a residual analysis and an analysis of variance are not compared for these different models. Future applications may find ways to overcome the statistical barriers presented by the degeneracy problem and the differences in ${\boldsymbol d}$ in order to test for which models are a better fit to the data.

The rest of this section discusses some of these issues. Section 5.2 discusses the relationship between the FM and the potential. Section 5.3 discusses how the accuracy of the other variables affects the results. A comparison with previous estimates is used as a starting point for discussing accuracy. Section 5.4 discusses continuity. This is relevant to determining whether the results for $V_{{r} \hskip 0.5pt 0}$ are reasonable. Section 5.5 discusses future applications of the method.

\subsection{Relationship between the FM and the Potential}

Analytical expressions for the relationship between the velocity perturbations and the potential, $\mathcal{P}$, in the presence of a weak bar-like perturbation are commonly derived in the literature by assuming that the first-order velocity perturbations are small and that there are only conservative forces at work (e.g., Binney \& Tremaine 2008, Chapter 3). These assumptions are problematic for constructing a complete picture of the dynamics of the oval distortion of NGC 4736. Derivations for the velocity perturbations that relate them to $\mathcal{P}$ are nontrivial when there are nonconservative forces at work.

The first-order velocity perturbations are only small near the end of the oval distortion. This is true whether or not there are net radial flows. Figure 11 shows the percent ratio of  $V_{{\theta} \hskip 0.5pt 2}$ to $V_{{\theta} \hskip 0.5pt 0}$ using the results for E+R and Eonly. Both models produce results that are relatively small near the end of the oval distortion, but the ratio increases to above 30$\%$ at the beginning of the oval distortion. The ratios for Eonly are larger than the ratios for E+R. 

Nonconservative forces must be at work in order for material to flow inward. In the presence of nonconservative forces, an analytical expression for the relationship between the velocities in the FM and $\mathcal{P}$ are obtainable from the Lagrangian equation of the second kind for generalized coordinates $q$, 
\begin{equation}
\frac{d}{dt} \frac{\partial \mathcal{L}}{\partial \dot{q}} - \frac{\partial \mathcal{L}}{\partial q} = Q_q, 
\end{equation}
where that $\mathcal{L}$ is the Lagrangian and $Q_q$ takes into account the work done by nonconservative forces (Joos \& Freeman 1986, Chapter 6). Considering only polar coordinates $r$ and $\theta$,
\begin{equation}
\mathcal{L} = \frac{1}{2}\big(\dot{r}^2 + r^2\dot{\theta}\big) - \mathcal{P}(r,\theta).
\end{equation}
Applying Equation (26) to Equation (27) leads to two coupled partial differential equations,
\begin{equation}
\frac{\partial \mathcal{P}}{\partial r} - Q_r = r\,\dot{\theta}^2 - \ddot{r}
\end{equation}
and
\begin{equation}
\frac{\partial \mathcal{P}}{\partial \theta} -Q_{\theta} = -2\,r\,\dot{r}\,\dot{\theta} - r^2\ddot{\theta}.
\end{equation}
Equation (28) describes the force per unit mass in the radial direction. Equation (29) describes the torque per unit mass about the origin. The functional forms of $Q_r$ and $Q_\theta$ are unknown, and this is why that derivations for the velocity perturbations that relate them to $\mathcal{P}$ are nontrivial. 

\begin{figure}[t!]
\includegraphics[width=0.99\columnwidth]{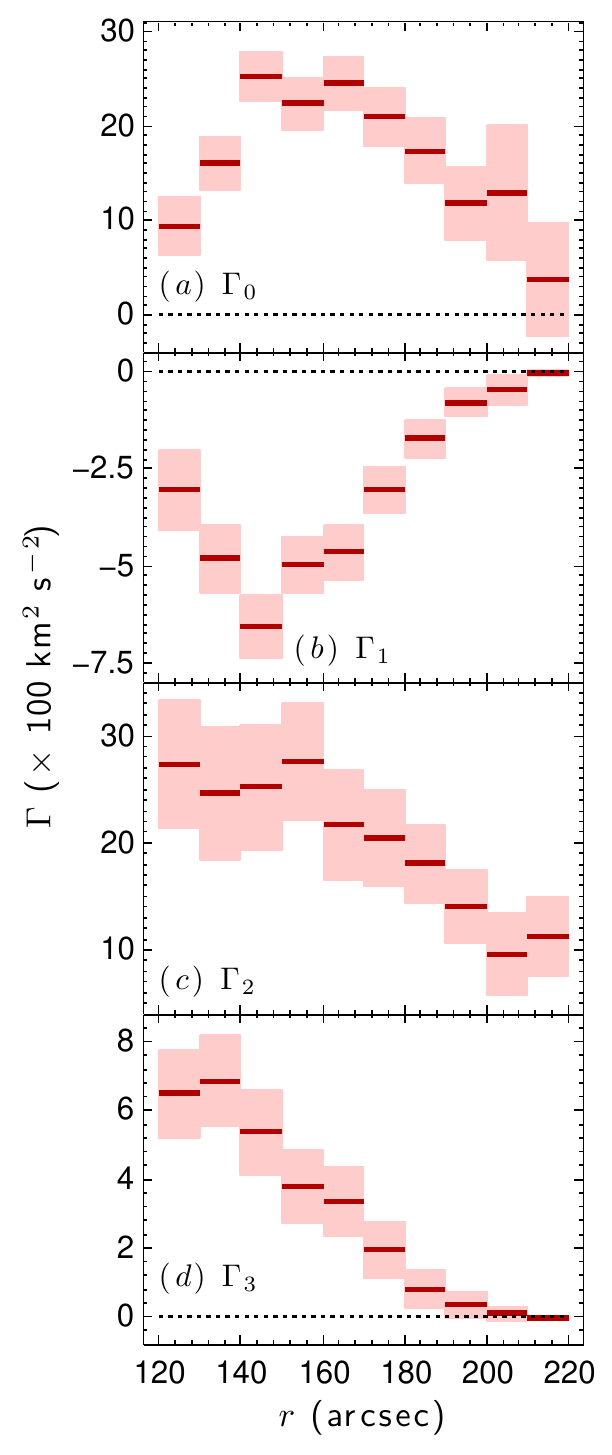} 
\caption{Radial profiles of $\Gamma_0$ -- $\Gamma_2$ calculated from the results in Section 4.4.  Each calculated coefficient is labeled in the bottom left of each panel. The dark-red line segments and light-red shading show the results and the 95\% CIs respectively.  The black dotted lines at 0 km$^2$ s$^{-2}$ in panels ($a$) and ($d$) are provided for reference.}
\end{figure}

For the FM and assumptions made in this paper, the time derivatives on the right-hand sides of Equations (28) and (29) are
\begin{equation}
\dot{r} = V_{r \hskip 0.5pt 0}(r)- V_{r \hskip 0.5pt 2}(r)\,\mbox{sin}(\Theta),
\end{equation}
%
\begin{equation}
\ddot{r} = -2\,(\dot{\theta} - \Omega_o)\,V_{r \hskip 0.5pt 2}(r)\,\mbox{cos}(\Theta),
\end{equation}
and
\begin{equation}
\ddot{\theta} =2\,(\dot{\theta} - \Omega_o)\,\frac{V_{\theta \hskip 0.5pt 2}(r)}{r}\,\mbox{sin}(\Theta)  - \frac{\dot{r}}{r}\,\dot{\theta},
\end{equation}
with $\dot{\theta}$ given by Equation (13). Substituting these into Equations (28) and (29), performing some algebra, and using a double angle formula, results in the following relationships:
\begin{align}
\frac{\partial \mathcal{P}}{\partial r} - Q_r = \hskip 2pt & F_0(r) + F_1(r) \,\mbox{cos}(\Theta) \nonumber \\ & + F_2(r)\,\mbox{cos}^2(\Theta)
\end{align}
and
\begin{align} 
\frac{\partial \mathcal{P}}{\partial \theta} -Q_{\theta} = \hskip 2pt &\Gamma_0(r)  + \Gamma_1(r) \,\mbox{cos}(\Theta) \nonumber \\ & +\Gamma_2(r)\,\mbox{sin}(\Theta) \nonumber \\ & + \Gamma_3(r)\,\mbox{sin}(2\Theta),
\end{align}
 where, for ease of notation, the coefficients are
\begin{align}
F_0(r) = \hskip 2pt &  \frac{V_{\theta \hskip 0.5pt 0}^2(r)}{r},\\
F_1(r) = \hskip 2pt & 2\,\Big\{\frac{V_{\theta \hskip 0.5pt 0}(r)}{r}\big[V_{r \hskip 0.5pt 2}(r) - V_{\theta \hskip 0.5pt 2}(r)\big] \nonumber \\  & - \Omega_o\,V_{r \hskip 0.5pt 2}(r)\Big\}, \\
F_2(r) = \hskip 2pt & \frac{V_{\theta \hskip 0.5pt 2}(r)}{r}\big[V_{\theta \hskip 0.5pt 2}(r) - 2\,V_{r \hskip 0.5pt 2}\big],\\
\Gamma_0(r) = \hskip 2pt & -V_{\theta \hskip 0.5pt 0}(r)V_{r \hskip 0.5pt 0}(r), \\
\Gamma_1(r) = \hskip 2pt & V_{r \hskip 0.5pt 0}(r)V_{\theta \hskip 0.5pt 2}(r), \\
\Gamma_2(r) = \hskip 2pt & V_{\theta \hskip 0.5pt 0}(r)\big[V_{r \hskip 0.5pt 2}(r) - 2\,V_{\theta \hskip 0.5pt 2}(r)\big] + 2\,r\,\Omega_o\,V_{\theta \hskip 0.5pt 2}(r),
\end{align}
and
\begin{equation}
\hskip -90pt \Gamma_3(r) = V_{\theta \hskip 0.5pt 2}^2(r) - \frac{1}{2}V_{\theta \hskip 0.5pt 2}(r)V_{r \hskip 0.5pt 2}(r).
\end{equation}
%

\begin{figure}[t!]
\includegraphics[width=1\columnwidth]{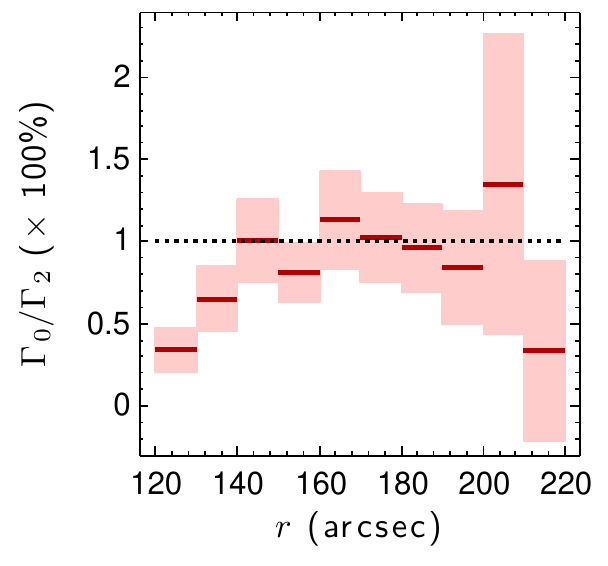} 
\caption{Percent ratio of $\Gamma_0$ to $\Gamma_2$.  The dark-red line segments and light-red shading show the results and the 95\% CIs, respectively.  The black dotted line at 100\% is provided for reference.}
\end{figure}
%
\begin{table*}[t!]
\centering
\caption{Previous Estimates of $\psi_{\mbox{\footnotesize inc}}$ and $\phi_{\mbox{\footnotesize maj}}$}
\begin{tabular}{cclll}
\hline
\hline
$\psi_{\mbox{\footnotesize inc}}$ & $\phi_{\mbox{\footnotesize maj}}$ & Method & Summary & Reference\\
(deg) & (deg) & & & \\
\hline
\hskip -22pt 42 & \hskip -16pt 288.7 & Kinematic & Fits circular velocities between the end of the oval and $R_{25}$. & This paper \\
\hskip -15 pt 41.4 & \hskip -16pt 296.1 & Kinematic  & Reports the mean from tilted rings for the whole disk. & (1) \\
$\cdot\cdot\cdot$ & \hskip -23.5pt 295 &  Kinematic & Notes that this $\phi_{\mbox{\footnotesize maj}}$ is the same as the pseudoring for their data, and & (2) \\
&&& sets inflow (outflow) inside (outside) the oval distortion for C+R. & \\
\hskip -22.7pt 35 & \hskip -23.5pt 290 & Photometric & Fits ellipses to isophotes between the inner pseudoring and the oval. & (3) \\
\hskip -4.5pt 30 $\pm$ 5 & \hskip -5.5pt 293 $\pm$ 1 & Kinematic & Fits circular velocities in the nuclear region. & (3) \\
\hskip -4.5pt 40 $\pm$ 5 & 290 - 310 & Kinematic & Fits tilted rings for the whole disk. & (4) \\
\hskip -22.7pt 36 & \hskip -23.5pt 285 & Photometric & Fits an ellipse to an isophote at $R_{\mbox{\footnotesize 25}}$. & (5) \\
35 $\pm$ 10 & \hskip -5.5pt 302 $\pm$ 3 & Kinematic & Fits circular velocities in the nuclear region. & (6) \\
 & \hskip -5.5pt 294 $\pm$ 6 & Kinematic & Fits circular velocities for 120$^{\prime\prime}$ $<$ $r$ $<$ 220$^{\prime\prime}$. & (6) \\
\hskip -4.9pt 36 $\pm$ 6 & \hskip -5.5pt 294 $\pm$ 2 & Kinematic & Fits circular velocities for $r$ $<$ 70$^{\prime\prime}$. & (7) \\
$\cdot\cdot\cdot$ & \hskip -23.5pt 303 & Kinematic & Finds a $\phi_{\mbox{\footnotesize maj}}$ that makes the inner pseudoring round for their data. & (8) \\[3pt]
\hline
\multicolumn{5}{l}{{\bf References}--(1) dB08; (2) WB00; (3) M95; (4) MVD93; (5) de Vaucoulers al. 1991; (6) Bosma et al. 1977;}\\
\multicolumn{5}{l}{(7) van der Kruit 1976; (8) Burbidge \& Burbidge (1962).}\\
\end{tabular}
\end{table*}

The net radial flow velocity, $V_{r \hskip 0.5pt 0}$, is absent in Equation (33), and present in Equation (34). Setting $Q_\theta$ = 0 in Equation (34) is clearly nonphysical due to the first term on the right-hand side of that equation, $\Gamma_0$, that contains $V_{r \hskip 0.5pt 0}$. Doing so requires a positive increasing trend in $\mathcal{P}$ as a function of $\theta$, but the physical location of $\theta$ repeats every 360$^\circ$.

The non-negligible effect of $Q_\theta$ is demonstrated in Figure 12. The $\Gamma_2$ coefficient in front of the sin($\Theta$) term in Equation (34) should dominate $\partial \mathcal{P}$/$\partial \mathcal{\theta}$ for the coordinate system and FM used in this paper. However, this is not the dominating coefficient for all of the fitted rings of data. Figure 13 shows the percent ratio of $\Gamma_0$ to $\Gamma_2$. The ratio is indistinguishable from 100\% for most of the results shown given the size of the 95\% CIs. The mean percent ratio for all of the results shown in Figure 13 is 84\% $\pm$ 23\%.

Any future derivations that relate $\mathcal{P}$ to the velocity perturbations of a weak bar-like potential will need to account for nonconservative forces where net radial flows are present, even if those flows are only on the order of 1 - 10 km s$^{-1}$. Although Equation (33) only contains velocities that are explainable in the absence of nonconservative forces, this does not guarantee that $Q_r$ = 0. The presence of $Q_\theta$ in Equation (34) may affect all of the velocities in that equation by some non-negligible amount, and all of the velocities in Equation (33) are also in Equation (34). For example, $V_{r \hskip 0.5pt 0}$ is directly coupled to $V_{\theta \hskip 0.5pt 0}$ and $V_{\theta \hskip 0.5pt 2}$ in the first and second terms on the right-hand side of Equation (34), respectively.

\subsection{Accuracy of the Other Variables}

\subsubsection{Accuracy of $\psi_{\mbox{\footnotesize inc}}$ and $\phi_{\mbox{\footnotesize maj}}$}

Table 3 summarizes the previous estimates of $\psi_{\mbox{\footnotesize inc}}$ and $\phi_{\mbox{\footnotesize maj}}$ for comparison with the values adopted in this paper. The differences in the previous estimates are greater than the reported uncertainties.  Previous estimates of $\psi_{\mbox{\footnotesize inc}}$ differ by as much as 11\fdg4, and those of $\phi_{\mbox{\footnotesize maj}}$ differ by as much as 25$^\circ$. These large differences are a consequence of different methods applied to different regions of this complex galaxy. Included in Table 3 is information about the methods used and short summaries about them.

Some of the previous estimates in Table 3 deserve further comment. The results from tilted rings consistently show the same variation in the radial profile of $\phi_{\mbox{\footnotesize maj}}$ (dB08; MVD93; Figure 2 of this paper). Ellipses fit to isophotes are unreliable for estimating $\psi_{\mbox{\footnotesize inc}}$ and $\phi_{\mbox{\footnotesize maj}}$ near the edge of the optical disk because the outer pseudoring of spiral arms overlaps with $R_{25}$ (e.g., Barnes \& Sellwood 2003). Fits for models that only include circular velocities are biased from excluding other velocity components when they are found for regions with well-defined patterns such as the nuclear region containing a bar and an inner pseudoring and the region containing the oval distortion. 

The value of $\psi_{\mbox{\footnotesize inc}}$ is better constrained than $\phi_{\mbox{\footnotesize maj}}$, especially when only considering the more reliable kinematic methods that include data beyond the nuclear region. The two previous methods that satisfy this criterion estimate values of 41\fdg4 (dB08) and 40$^\circ$ $\pm$ 5 (MVD93). These are consistent with the adopted value given the uncertainty reported by MVD93, and the typical uncertainties reported in the literature for the type of high-quality data in this paper.

\begin{figure}[t!]
\includegraphics[width=1\columnwidth]{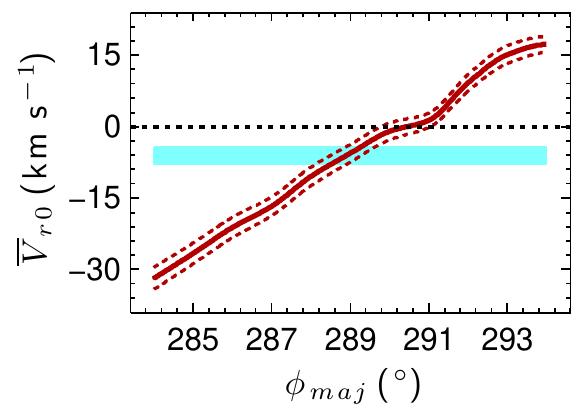} 
\caption{Dependence of the $\overline{V}_{{r} \hskip 0.5pt 0}$ results from E+R on $\phi_{\mbox{\footnotesize maj}}$. The red solid and dotted lines show the results for $\overline{V}_{{r} \hskip 0.5pt 0}$ and the 95\% CIs, respectively. The cyan shading shows the 95\% CI for $\overline{V}_{{r} \hskip 0.5pt 0}$ calculated from the results in Section 4.4 for the adopted value of $\phi_{\mbox{\footnotesize maj}}$. The black dotted line at 0 km s$^{-1}$ is provided for reference.}
\end{figure}
%
\begin{figure}[ht!]
\includegraphics[width=1\columnwidth]{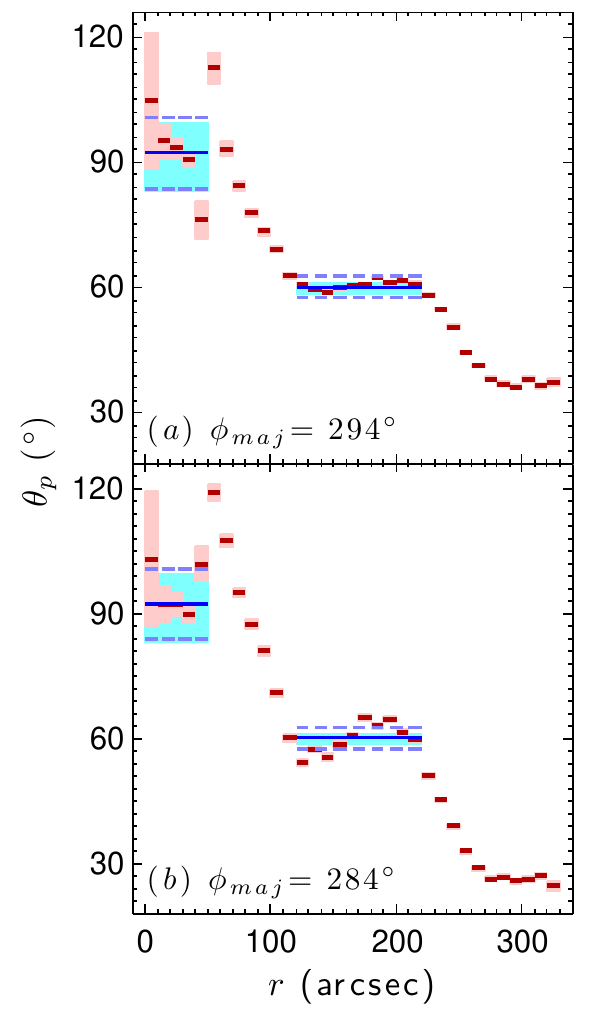} 
\caption{Results for calculations of Equation (17) for $\phi_{\mbox{\footnotesize maj}}$ = 284$^{\circ}$ and 294$^{\circ}$. The values of $\phi_{\mbox{\footnotesize maj}}$ are shown in the bottom left of each panel. The panels are formatted in the same way as in Figure 6. The cyan shading shows the 95\% CIs for the adopted values of $\theta_o$ and $\theta_b$.}
\end{figure}
%
\begin{figure}[ht!]
\includegraphics[width=1\columnwidth]{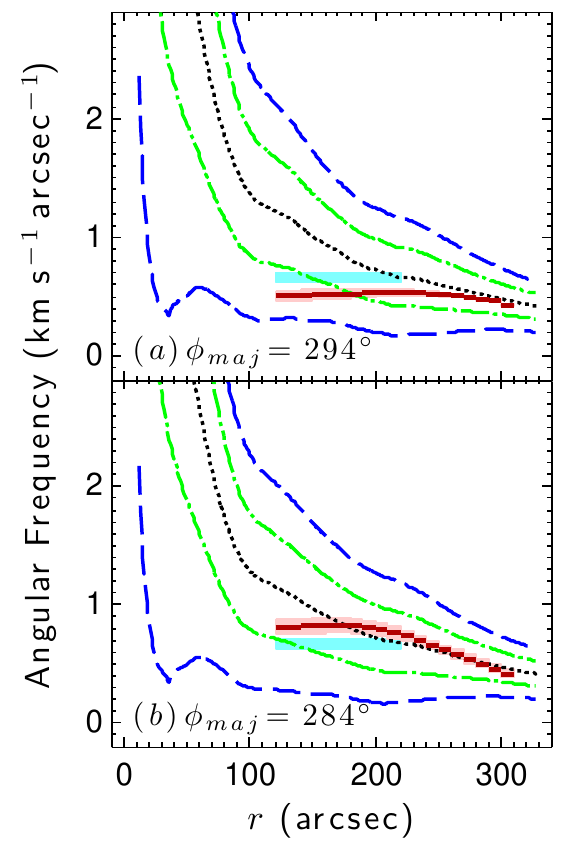} 
\caption{Results for the general form of the TW84 method that excludes integrals in the region $|y_i|$ $<$ 120$^{\prime\prime}$ for $\phi_{\mbox{\footnotesize maj}}$ = 284$^{\circ}$ and 294$^{\circ}$. The values of $\phi_{\mbox{\footnotesize maj}}$ are shown in the bottom left of each panel. The panels are formatted in the same way as panel ($b$) in Figure 7. The possible locations for resonance are found using the value of $\phi_{\mbox{\footnotesize maj}}$ shown in the bottom left of each panel. The cyan shading shows the 95\% CI for the adopted value of $\Omega_o$.}
\end{figure}

\begin{figure}[t!]
\includegraphics[width=1\columnwidth]{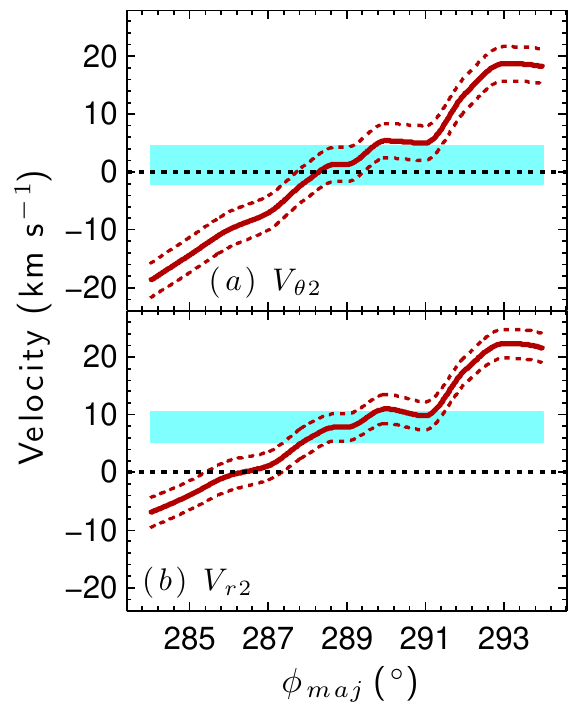} 
\caption{ Dependance of the $V_{\theta \hskip 0.5pt 2}$ and $V_{r \hskip 0.5pt 2}$ results from E+R on $\phi_{\mbox{\footnotesize maj}}$ for the outermost ring of data in the oval distortion. The different velocity components are labelled in each panel. The red solid and dotted lines show the results and the 95\% CIs, respectively. The cyan shading shows the 95\% CIs for their respective results in Section 4.4 that are found for the adopted value of $\phi_{\mbox{\footnotesize maj}}$. The black dotted lines at 0 km s$^{-1}$ are provided for reference.}
\end{figure}

\begin{figure*}[t!]
\includegraphics[width=1\textwidth]{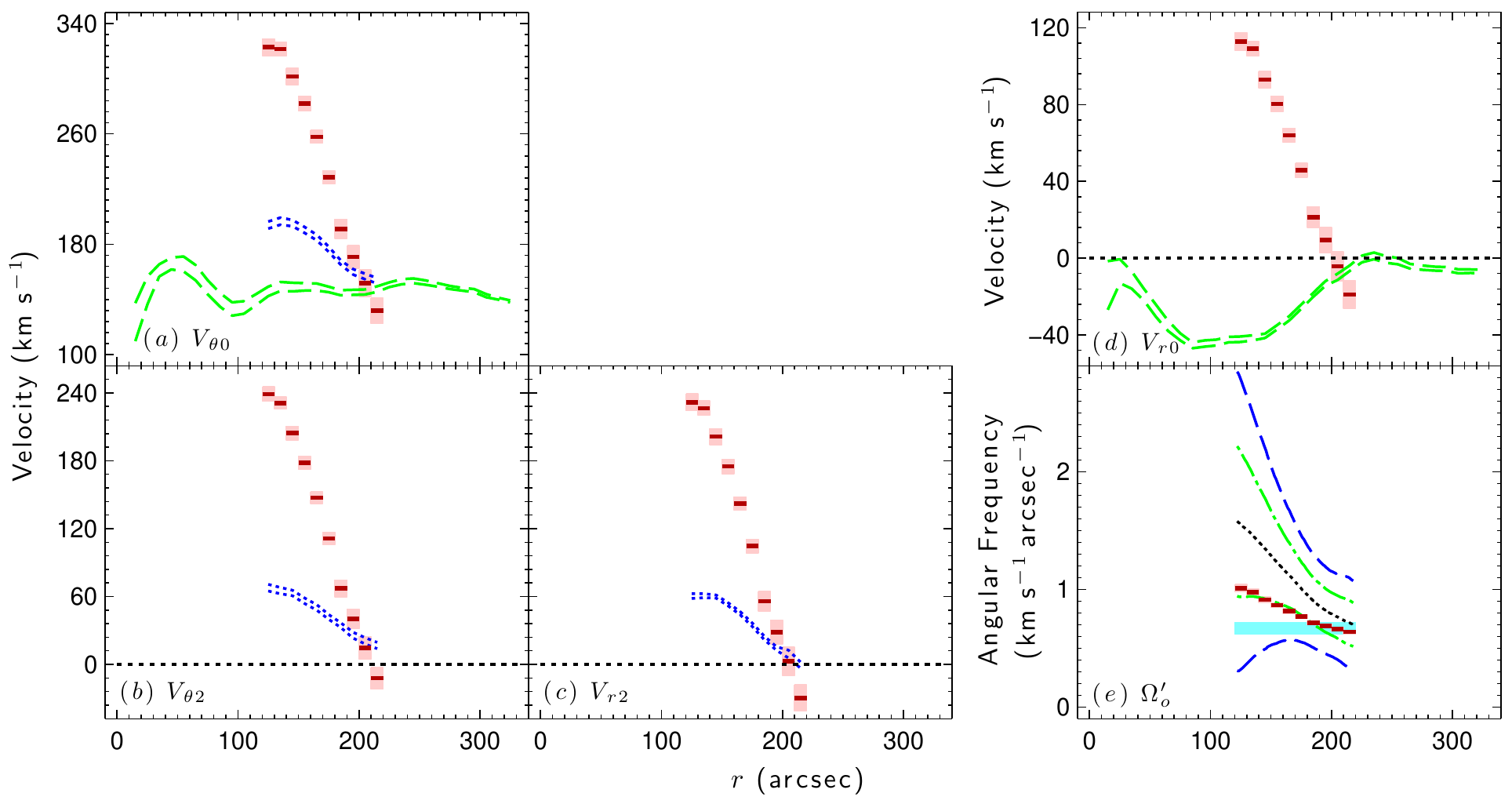} 
\caption{Results for modified and reduced velocity models for $\theta_o$ = 20\fdg6. The figure is formatted in the same way as Figure 10.}
\end{figure*}

The value of $\phi_{\mbox{\footnotesize maj}}$ is more challenging to estimate owing to the possible bias from excluding radial velocity components. Conversely, a change in $\phi_{\mbox{\footnotesize maj}}$  affects the measured values of $V_{{r} \hskip 0.5pt 0}$. Figure 14 shows $\overline{V}_{{r} \hskip 0.5pt 0}$  results from E+R for 284$^\circ$ $\leqslant$ $\phi_{\mbox{\footnotesize maj}}$ $\leqslant$ 294$^\circ$.  The results for $\overline{V}_{{r} \hskip 0.5pt 0}$ increase with increasing $\phi_{\mbox{\footnotesize maj}}$. The value of $\overline{V}_{{r} \hskip 0.5pt 0}$ = 0 km s$^{-1}$ when $\phi_{\mbox{\footnotesize maj}}$ = 290\fdg1, which is 1\fdg4 larger than the adopted value of $\phi_{\mbox{\footnotesize maj}}$. This is within the typical uncertainty of a few degrees that is reported in the literature for the type of high-quality data in this paper, but it is 1$^\circ$ more than the uncertainty of 0\fdg4 for the mean of the results using tilted rings in the region between the end of the oval distortion and $R_{25}$. Values of $\phi_{\mbox{\footnotesize maj}}$ $>$ 290\fdg1 produce net outflows, inconsistent with expectations. 

The sensitivity of the $V_{{r} \hskip 0.5pt 0}$ results to an accurate estimate of $\phi_{\mbox{\footnotesize maj}}$ necessitates more justification for the adopted value than minimizing the SSRs over the region showing an approximately constant radial profile for $\phi_{\mbox{\footnotesize maj}}$. The entire process of measuring $\theta_o$, $\Omega_o$, and the velocities is investigated for different $\phi_{\mbox{\footnotesize maj}}$ to see what other reasons there are for the adopted value. For $\theta_o$ and $\Omega_o$, only the results for $\phi_{\mbox{\footnotesize maj}}$ = 284$^\circ$ and 294$^\circ$ are shown for brevity, but they are sufficient for this purpose.

Figure 15 shows the results for $\theta_p$ when $\phi_{\mbox{\footnotesize maj}}$ = 284$^\circ$ and 294$^\circ$. Changing $\phi_{\mbox{\footnotesize maj}}$ to these values has very little effect on the estimates of $\theta_o$. In the figure one can see that $\theta_o$ for the different $\phi_{\mbox{\footnotesize maj}}$ are consistent with the adopted value of $\theta_o$ given the size of the 95\% CIs.

Figure 16 shows the results for $\Omega_p$ when $\phi_{\mbox{\footnotesize maj}}$ = 284$^\circ$ and 294$^\circ$. The effect on $\Omega_p$ is much greater than the 95\% CIs. Increasing $\phi_{\mbox{\footnotesize maj}}$ decreases $\Omega_p$, and extends the radius of the corotation resonance beyond the oval distortion. Decreasing $\phi_{\mbox{\footnotesize maj}}$ increases $\Omega_p$, and decreases the radius of the corotation resonance, placing it within the oval distortion. Oval distortions consisting of $x_1$-type orbits are expected to extend approximately up to corotation, but not beyond it (C80; Tueben et al. 1986). The results in Figure 16 therefore argue against values of $\phi_{\mbox{\footnotesize maj}}$ that are lower than the adopted value.

The effect different values of $\phi_{\mbox{\footnotesize maj}}$ have on the amplitudes of the velocity perturbations induced by the oval distortion is parameterized using the $V_{{\theta} \hskip 0.5pt 2}$ and $V_{{r} \hskip 0.5pt 2}$ results for the outermost ring of the oval distortion. The radial trends in $V_{{\theta} \hskip 0.5pt 2}$ and $V_{{r} \hskip 0.5pt 2}$ should be positive in sign and approach zero near the end of the oval distortion owing to the proximity of a corotation resonance beyond the oval distortion. Figure 17 shows the results from E+R for 284$^\circ$ $\leqslant$ $\phi_{\mbox{\footnotesize maj}}$ $\leqslant$ 294$^\circ$.  Both $V_{{\theta} \hskip 0.5pt 2}$ and $V_{{r} \hskip 0.5pt 2}$ increase with increasing $\phi_{\mbox{\footnotesize maj}}$. At the adopted value of $\phi_{\mbox{\footnotesize maj}}$ = 288\fdg7, $V_{{\theta} \hskip 0.5pt 2}$ is consistent with 0 km s$^{-1}$ given the size of the 95\% CIs. The values of $V_{{r} \hskip 0.5pt 2}$ are consistent with 0 km s$^{-1}$ for 285\fdg3 $\leqslant$ $\phi_{\mbox{\footnotesize maj}}$ $\leqslant$ 287\fdg4 given the size of the 95\% CIs. However, values of $\phi_{\mbox{\footnotesize maj}}$ that are this small result in negative values for $V_{{\theta} \hskip 0.5pt 2}$. They also increase $\Omega_o$ relative to $\Omega$, placing the corotation resonance well within the oval distortion. As pointed out for panel ($b$) in Figure 16, a corotation resonance well within the oval distortion is inconsistent with the assumption of $x_1$-type orbits.

There are two conclusions from exploring different values of $\phi_{\mbox{\footnotesize maj}}$ to see what other reasons there are for the adopted value. The first is that the results for $\Omega_o$ and $V_{{r} \hskip 0.5pt 0}$ are very sensitive to the accuracy of $\phi_{\mbox{\footnotesize maj}}$. The second is that the adopted value must be very close to the true value on the grounds that the $\Omega_p$ and velocity results are more consistent with expectations than the results obtained for most of the different $\phi_{\mbox{\footnotesize maj}}$. The second conclusion is established from values of $\overline{V}_{{r} \hskip 0.5pt 0}$ for $\phi_{\mbox{\footnotesize maj}}$ $>$ 290\fdg1 in Figure 14 that are inconsistent with expectations, the results for $\Omega_p$ in Figure 16 arguing against $\phi_{\mbox{\footnotesize maj}}$ $<$ 288\fdg7, and the increasing values of $V_{\theta \hskip 0.5pt 2}$ and $V_{r \hskip 0.5pt 2}$ in Figure 17 for $\phi_{\mbox{\footnotesize maj}}$ $>$ 288\fdg7.

\subsubsection{Accuracy of $\theta_o$}

Previous estimates for the location of the oval distortion are performed in sky coordinates by fitting ellipses to isophotes. Standard procedures for these fits are based on the method of Jedrzejewski (1987, and references therein; hereafter J87). The method begins with initial estimates of the center, ellipticity, $\epsilon$, and position angle from north to east in the sky, $\phi_o$. The data are then sampled along this initial ellipse, and a model of the form
\begin{align}
I  =  I_0 \,\,+  & \,\,A_1\,\mbox{sin}(\phi) + B_1\,\mbox{cos}(\phi) \nonumber \\ & \hskip -10pt + A_2\,\mbox{sin}(2\phi) +  B_2\,\mbox{cos}(2\phi)
\end{align}
is fit to the intensity, $I$, of the sampled data. The coefficients in front of the sine and cosine terms describe the deviations from a true ellipse and contain information for updating the initial values. The initial values are updated, and the process is repeated until the coefficients are less than the rms of the residuals by an arbitrary amount that is determined during implementation. The relationship between the coefficients and the updates for the initial values are provided in J87. Previous estimates of $\phi_o$ using this method are 95$^{\circ}$ (M95), 90$^{\circ}$ (Erwin 2004, hereafter E04), and 94\fdg4 (Comer\'{o}n et al. 2014). The mean of the previous estimates is 93\fdg1 $\pm$ 6\fdg6. 

Azimuthal sky coordinates, $\phi$, are related to galaxy coordinates, $\theta$, according to
\begin{equation}
\mbox{tan}(\theta)\,\mbox{cos}(\psi_{\mbox{\footnotesize inc}}) = \mbox{tan}(\phi_{\mbox{\footnotesize maj}} - \phi).
\end{equation}
Equation (43) is derivable using the definitions tan($\phi$) = ($\alpha - \alpha_{\mbox{\footnotesize kc}}$)/($\delta - \delta_{\mbox{\footnotesize kc}}$), tan($\theta$) = $y$/$x$, and the difference formula for tangent see also VdKA78, but note that the sign difference for their result is due to the sign convention used in this paper for defining $y$ in Section 3.1). The value of 93\fdg1 in sky coordinates converts to 20\fdg6 in galaxy coordinates. This is 39\fdg2 less than the adopted value of $\theta_o$ = 59\fdg8 $\pm$ 1\fdg4. The adopted value of $\theta_o$ converts to $\phi_o$ = 56\fdg8 $\pm$ 1\fdg6 in sky coordinates.

The results for the velocity models using $\theta_o$ = 20\fdg6 are shown in Figure 18. Most of the results for $V_{{\theta} \hskip 0.5pt 0}$, $V_{{\theta} \hskip 0.5pt 2}$, and $V_{{r} \hskip 0.5pt 2}$ for E+R and Eonly are much larger than what is shown in Figure 10. The mean of $V_{{r} \hskip 0.5pt 0}$ is 66.1 $\pm$ 34.2 km s$^{-1}$. The calculated results for $\Omega_o^\prime$ show a radially decreasing trend along the inner ultraharmonic (4:1) Lindblad resonance that is inconsistent with an approximately rigidly rotating oval. A value of $\theta_o$ = 20\fdg6 clearly produces nonphysical results.

\begin{figure*}[t!]
\includegraphics[width=1\textwidth]{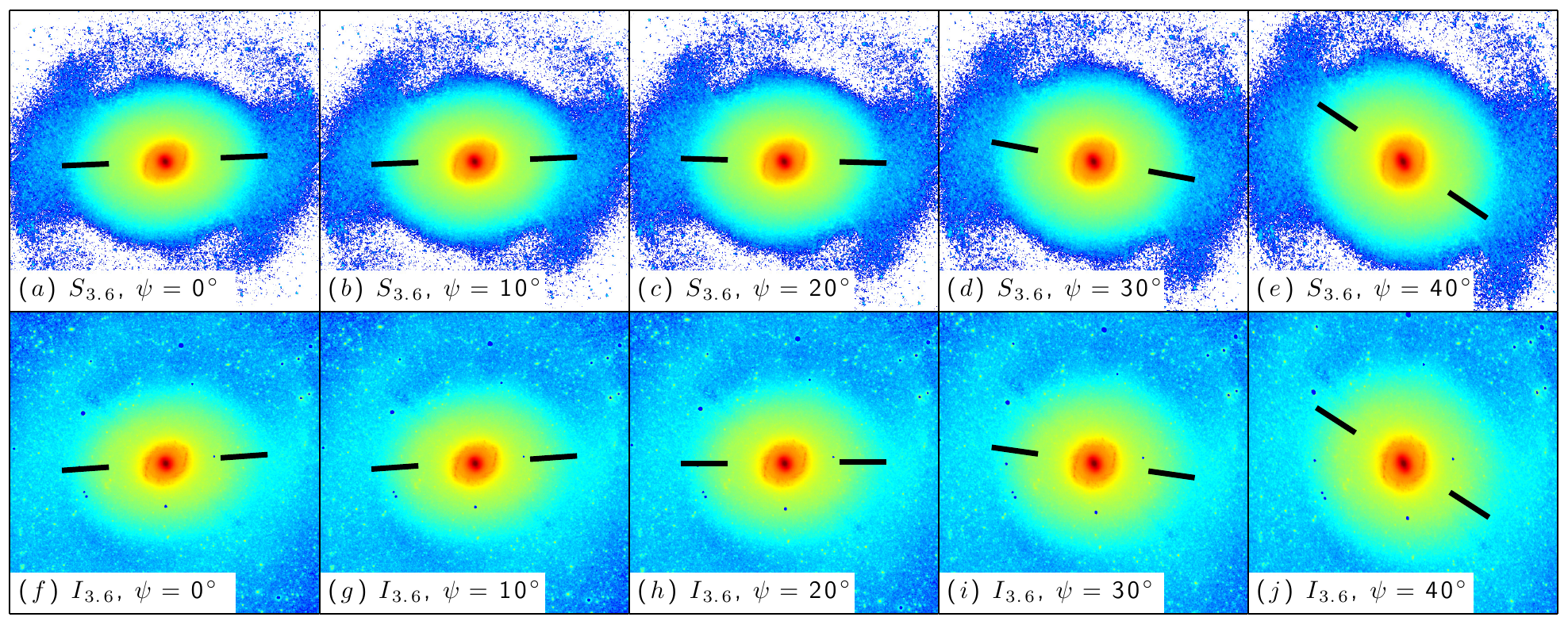} 
\caption{Variations in the shape of the oval distortion as the disk is deprojected along the minor kinematic axis. The data type and the deprojection angle are shown in the lower left corner of each panel. Both data types are log scaled. The black solid line segments show the major axis and extend from the beginning to the end of the oval distortion. The horizontal and vertical axes in panels ($a$) and ($f$) are parallel to those in Figure 1.}
\end{figure*}

The value of $\theta_o$ = 20\fdg6, however, is calculated from three different measurements, and fitting ellipses to isophotes is a well-established procedure. Three experiments are therefore performed that rule out the explanation that the difference between previous estimates and the adopted value of $\theta_o$ is due to the method or data used. The details of the experiments and their results are presented in Appendix B. 

The difference is explainable by noting that the true shape and orientation of the oval distortion are best measured in galaxy coordinates owing to the small ellipticity  of the oval distortion. The ellipticity of the oval distortion { in sky coordinates, $\epsilon$} = 0.23 (M95), is close to the practical limit for reliable measurements of $\phi_o$ (e.g., Busko 1996). For such a small ellipticity, even small deviations from a true ellipse can greatly affect its projected shape. In sky coordinates M95 show that the oval distortion is cuspy. In galaxy coordinates, however, it appears more boxy. Two figures are provided to demonstrate this explanation. 

Figure 19 demonstrates how the cuspy shape of the oval distortion changes to a more boxy one as the galaxy is deprojected in inclination. The galaxy is deprojected from $\psi$ = 0$^\circ$ to 40$^\circ$ along the kinematic minor axis in 10$^\circ$ increments. The results for both $S_{3.6}$ and $I_{3.6}$ are shown for comparison. The position angles of the oval distortion shown in the figure are found using this paper's method in the deprojected sky coordinates. 

\begin{figure}[t!]
\center
\includegraphics[width=1\columnwidth]{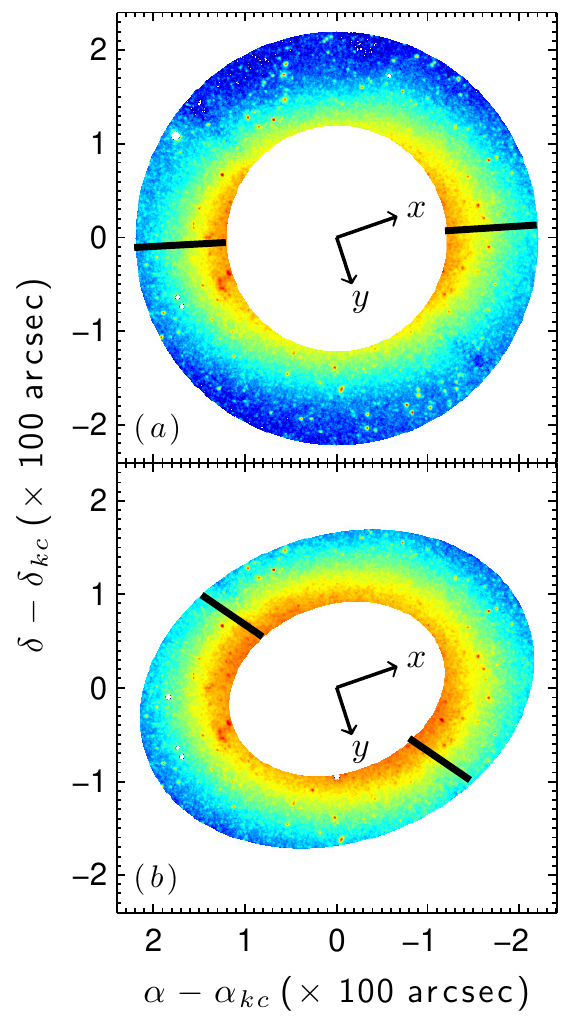} 
\caption{Annuli of $I_{3.6}$ showing how the apparent orientation of the oval distortion in sky coordinates is different from the true orientation in galaxy coordinates. The data are log scaled. Panels ($a$) and ($b$) show annuli from 120$^{\prime\prime}$ to 220$^{\prime\prime}$ in sky and galaxy coordinates, respectively (see text for details). The black arrows in each panel show the directions of $x$ and $y$ in galaxy coordinates. The black solid line segments showing the apparent major axes of the oval distortion are aligned with $\theta_o$ = 20\fdg6 ($\phi_o$ = 93\fdg1) and $\theta_o$ = 59\fdg8 ($\phi_o$ = 56\fdg8) for panels ($a$) and ($b$), respectively. }
\end{figure}
%
\begin{figure}[t!]
\center
\includegraphics[width=1\columnwidth]{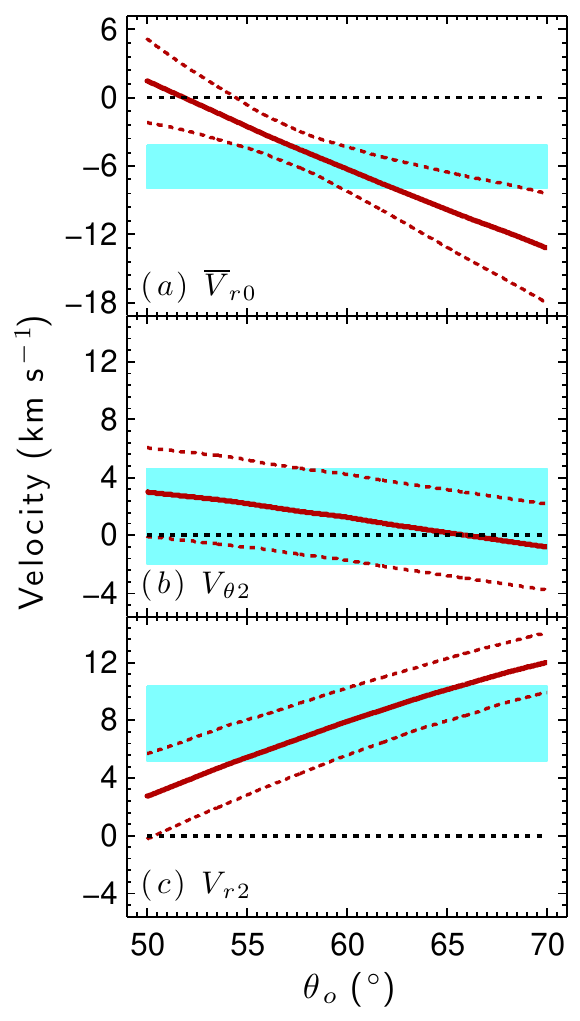} 
\caption{Dependence of the noncircular velocity results from E+R on $\theta_o$. The different velocity components are labeled in each panel. The $V_{\theta \hskip 0.5pt 2}$ and $V_{r \hskip 0.5pt 2}$ results are for the outermost ring of data in the oval distortion. The cyan shading shows the 95\% CIs for their respective results in Section 4.4 that are found for the adopted value of $\theta_o$. The black dotted lines at 0 km s$^{-1}$ are provided for reference.}
\end{figure}

Figure 20 demonstrates how the orientation of the oval distortion appears different in sky and galaxy coordinates by showing different annuli of $I_{3.6}$. Panel ($a$) only includes data for 120$^{\prime\prime}$ $\leqslant$ $R$ $\leqslant$ 220$^{\prime\prime}$, where that,
\begin{equation}
R = \sqrt{(\alpha - \alpha_{\mbox{\footnotesize kc}})^2 + (\delta - \delta_{\mbox{\footnotesize kc}})^2}.
\end{equation}
Panel ($b$) only includes data for 120$^{\prime\prime}$ $\leqslant$ $r$ $\leqslant$ 220$^{\prime\prime}$.  The major axis in panel ($a$) aligns well with the apparent orientation of the oval distortion in that panel. The orientation of the oval distortion is clearly different in panel ($b$). The orientation of the oval distortion in panel ($b$) is consistent with the results shown in Figure 5 for that region.

Unlike $\theta_o$, the measurement of $\theta_b$ is unaffected by the coordinate system used. The previous estimates of $\phi_b$ converted to galaxy coordinates are 87\fdg1 (M95) and 89\fdg6 (E04). Both of these estimates are from fitting ellipses to isophotes. The adopted value of $\theta_b$ = 91\fdg7 $\pm$ 8\fdg3 is consistent with these previous estimates given the size of the 95\% CI. Note that the bar ellipticity is 0.53 (M95), which is more than twice the ellipticity of 0.23 for the oval distortion. The consistency in the results for the bar supports the explanation that the differences between the adopted value of $\theta_o$ and the previous estimates are due to the small ellipticity of the oval distortion in combination with small deviations from a true ellipse and the effect these properties have on its projected shape in sky coordinates. 

Smaller changes in $\theta_o$ have a more modest effect on the results for the noncircular velocities than those shown in Figure 18. Figure 21 shows the results for the noncircular velocities from E+R for 50$^\circ$ $\leqslant$ $\theta_o$ $\leqslant$ 70$^\circ$. The $V_{\theta \hskip 0.5pt 2}$ and $V_{r \hskip 0.5pt 2}$ results are for the outermost ring of data in the oval distortion. The results for $\overline{V}_{{r} \hskip 0.5pt 0}$ and $V_{r \hskip 0.5pt 2}$ decrease, and the results for $V_{\theta \hskip 0.5pt 2}$ increase, with increasing $\theta_o$. The value of $\overline{V}_{{r} \hskip 0.5pt 0}$ = 0 km s$^{-1}$ when $\theta_o$ = 51\fdg9, which is 7\fdg9 less than the adopted value of $\theta_o$. The $V_{\theta \hskip 0.5pt 2}$ and $V_{r \hskip 0.5pt 2}$ results for the range of $\theta_o$ shown, however, are consistent with the results for the adopted value of $\theta_o$ = 59\fdg8 given the size of the 95\% CIs, thus providing no compelling reason for a better estimate of $\theta_o$.

\subsubsection{Accuracy of $\Omega_o$}

Previous estimates of $\Omega_p$ are summarized in Table 4 for comparison with $\Omega_o$ = $\Omega_p$ in the region of the oval distortion. The table shows estimates for different regions because the previous estimates of $\Omega_p$ are often assumed to be the same for multiple regions. They are adjusted for their adopted values of $\psi_{\mbox{\footnotesize inc}}$, and for their adopted distances if relevant.

Included in Table 4 is information about the methods used and short summaries about them. All of the previous estimates assume that $\Omega_p$ is a constant function of radius, but two of them (M95, Mu{\~n}oz-Tu{\~n}{\'o}n et al. 2004) allow for more than one value of $\Omega_p$. Most of the previous estimates interpret data in the context of theoretical model assumptions about how the possible locations of resonance coincide with photometric features such as the pseudorings and the beginnings and endings of the nuclear bar and oval distortion. Two of the previous estimates match simulations to observations. One previous estimate uses the kinematic method of TW84. 
\begin{table*}
\centering
\caption{Previous Estimates of $\Omega_p$}
\begin{tabular}{cllll}
\hline
\hline
$\Omega_p$ & Region & Method & Summary & Reference\\
(km s$^{-1}$ arcsec$^{-1}$) &  & & & \\
\hline
 0.67 $\pm$ 0.05 & Oval  & TW84 &  Applies the general method to H{\footnotesize I}. & This paper \\
 2.96 $\pm$ 0.36 &  Nuclear Bar & TW84 & Applies the general method to H{\footnotesize I}.  &  \\
 \hskip -30pt 0.86 & Whole Disk & Simulation  & Reproduces the inner and outer pseudorings. & (1)\\
 \hskip -30pt 2.33  & Nuclear Bar  & Interpretation  & Sets the outer Lindblad resonance of the bar at the   &  (2) \\
&  &  & inner Lindblad resonance of the bulge. &   \\
 2.80 $\pm$ 0.51 & Nuclear Bar & TW84 & Applies the original method to carbon monoxide.  &  (3) \\
 \hskip -30pt 0.76 & Whole Disk & Interpretation  & Sets corotation at the end of the oval.  &  (4) \\
 \hskip -30pt 0.79 & Whole Disk & Interpretation  & Sets corotation to where the inflow changes to outflow &  (5) \\
&  &  & for their C+R model results.  &  \\
 \hskip -30pt 1.22 & Whole Disk  & Simulation  & Reproduces the gas morphology.  & (6) \\
\hskip -30pt  8.42  & Nuclear Bar  & Interpretation  & Sets the outer Lindblad resonance of the bar at the &  (7)\\  
& & & inner pseudoring.  &  \\
\hskip -30pt  1.22 & Oval  & Interpretation & Sets the inner Lindblad resonance of the oval at the   &  (8)\\
& & & inner pseudoring.  &  \\
 0.92 $\pm$ 0.23 & Whole Disk & Interpretation  & Sets corotation at the gap in the optical intensity.   &  (9) \\[1pt]
\hline
\multicolumn{5}{l}{{\bf References}--(1) T09, (2) Mu{\~n}oz-Tu{\~n}{\'o}n et al. 2004, (3) RW04, (4) Waller et al. 2001, (5) WB00,   }\\
\multicolumn{5}{l}{(6) Mulder \& Combes 1996, (7) M95, (8) Buta 1988, (9) Schommer \& Sullivan 1976. }
\end{tabular}
\end{table*}
%

\begin{figure}[t!]
\includegraphics[width=1\columnwidth]{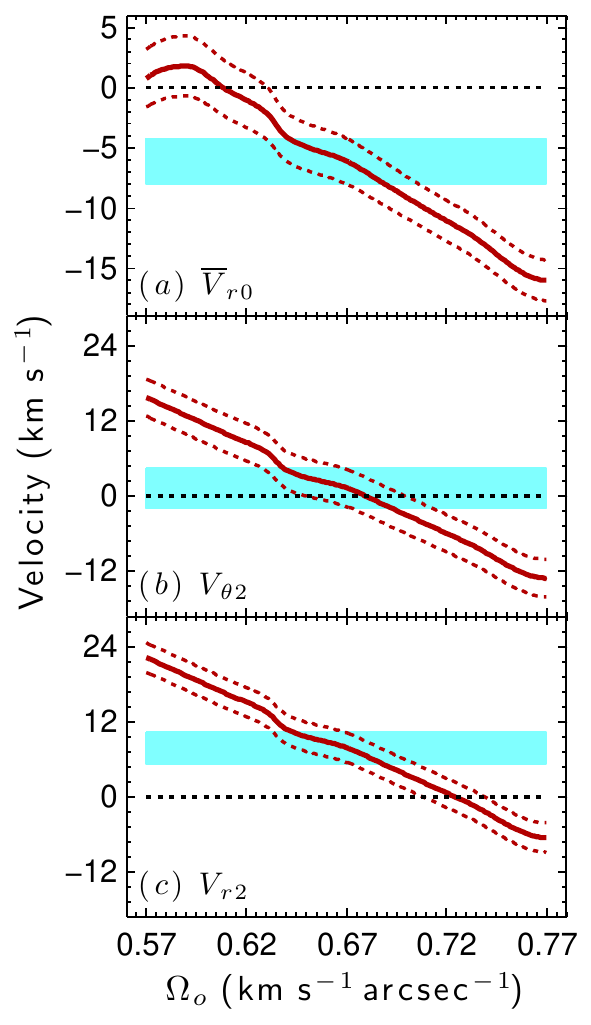} 
\caption{Dependence of the noncircular velocity results from E+R on $\Omega_o$. The panels of the figure are formatted in the same way as those in Figure 21.}
\end{figure}
%

\begin{figure}[ht!]
\includegraphics[width=1\columnwidth]{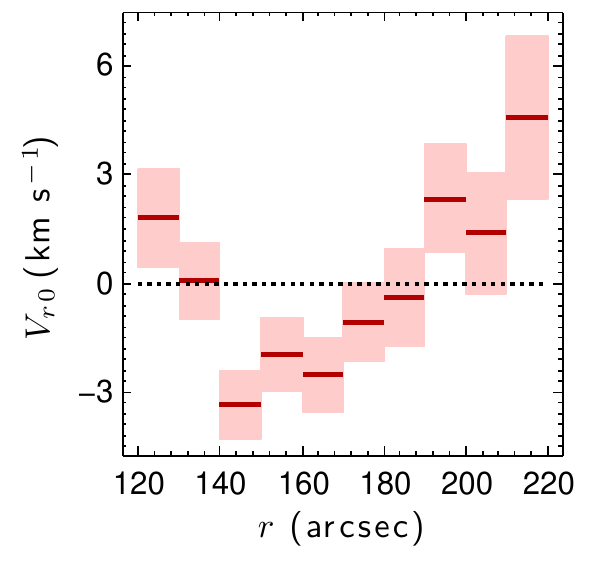} 
\caption{Results for $\overline{V}_{{r} \hskip 0.5pt 0}$ when $\phi_{\mbox{\footnotesize maj}}$ = 290\fdg1. The dark-red solid line segments and light-red shading show the results for each ring of data and the 95\% CIs, respectively. The black dotted line at 0 km s$^{-1}$ is provided for reference.}
\end{figure}
%

\begin{figure}[ht!]
\includegraphics[width=0.99\columnwidth]{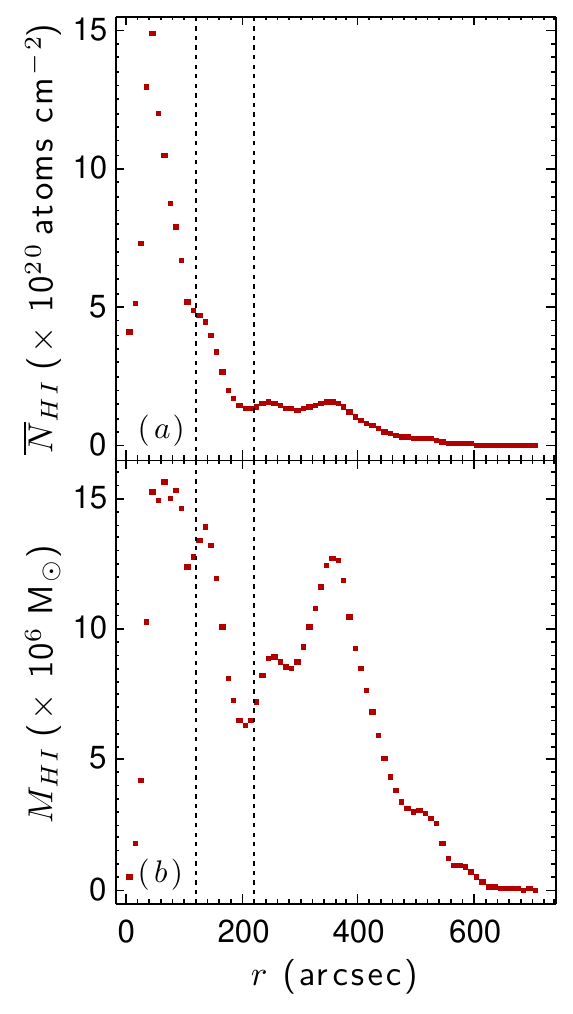} 
\caption{Radial profiles of $\overline{N}_{HI}$ and total $M_{HI}$. The vertical black dotted lines enclose the region of the oval distortion in both panels.}
\end{figure}
%

\begin{figure}[ht!]
\includegraphics[width=0.99\columnwidth]{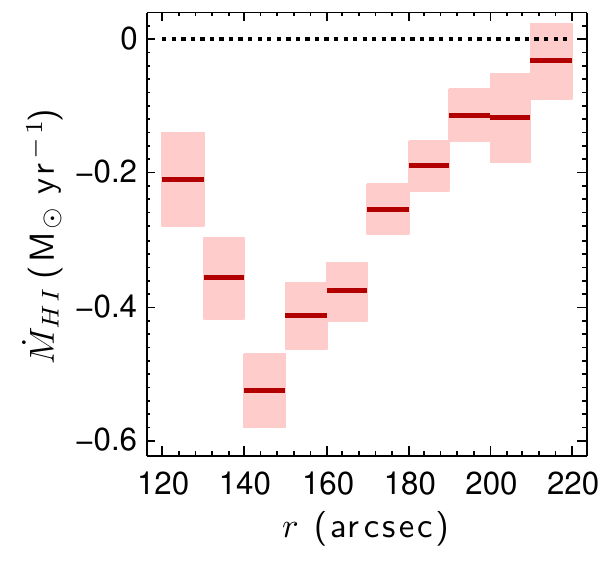} 
\caption{Estimates for the mass flow rate of H{\footnotesize I} in the oval distortion. The dark-red solid line segments and light-red shading show the results for each ring of data, and the 95\% CIs, respectively. The black dotted line at 0 km s$^{-1}$ is provided for reference.}
\end{figure}

The literature lacks a consensus for guiding a discussion about the accuracy of $\Omega_o$. This is a consequence of the diversity of the methods and assumptions. Further complicating any comparisons with the previous estimates is the need to adjust for differences in $\Omega_p$ that are due to differences in the adopted values of $\phi_{\mbox{\footnotesize maj}}$, which is nontrivial (e.g., Figure 16). Although a detailed comparison to all of the previous estimates of $\Omega_p$ is useful for evaluating the different theoretical model assumptions (see Dobbs \& Baba 2014 for a review), doing so is beyond the scope of this paper. It is worth pointing out, however, that the results in Figure 9 show that the oval distortion extends up to a corotation resonance as predicted by theory (C80, Tueben et al. 1986). 

This paper is the first to directly measure $\Omega_o$, and the second to directly measure $\Omega_p$ using the TW84 method. The previous estimate of $\Omega_p$ that is the most relevant for comparing to the results in this paper is the measurement by RW04. They measure $\Omega_b$ = $\Omega_p$ using carbon monoxide as a pattern tracer. After adjusting for their different $\psi_{\mbox{\footnotesize inc}}$, and converting to sky coordinates using their adopted distance, their value of $\Omega_b$ = 2.80 $\pm$ 0.51 km s$^{-1}$ arcsec$^{-1}$ is consistent with the value of $\Omega_b$ = 2.96 $\pm$ 0.36 found in this paper given the size of the 95\% CIs. 

Different values of $\Omega_o$ produce results for the noncircular velocities that are less consistent with expectations. Figure 22 shows the results for the noncircular velocities from E+R for 0.57 km s$^{-1}$ arcsec$^{-1}$ $\leqslant$ $\Omega_o$ $\leqslant$ 0.77 km s$^{-1}$ arcsec$^{-1}$. The $V_{\theta \hskip 0.5pt 2}$ and $V_{r \hskip 0.5pt 2}$ results are for the outermost ring of data in the oval distortion. The results for all three velocities shown in the figure decrease with increasing $\Omega_o$. The value of $\overline{V}_{{r} \hskip 0.5pt 0}$ = 0 km s$^{-1}$ when $\Omega_o$ = 0.61 km s$^{-1}$ arcsec$^{-1}$, which is 0.06 km s$^{-1}$ arcsec$^{-1}$ less than the adopted value of $\Omega_o$. Values of $\Omega_o$ less than the adopted value of 0.67 km s$^{-1}$ arcsec$^{-1}$ produce results for $V_{\theta \hskip 0.5pt 2}$ and $V_{r \hskip 0.5pt 2}$ that are larger than the results for the adopted value of $\Omega_o$. The values of $V_{{r} \hskip 0.5pt 2}$ are consistent with 0 km s$^{-1}$ for 0.71 km s$^{-1}$ arcsec$^{-1}$ $\leqslant$ $\Omega_o$ $\leqslant$ 0.74 km s$^{-1}$ arcsec$^{-1}$ given the size of the 95\% CIs. However, values of $\Omega_o$ that are this large result in values for $\overline{V}_{{r} \hskip 0.5pt 0}$ $<$ -10 km s$^{-1}$, and place the corotation resonance well within the oval distortion.

\subsection{Continuity}

The measured value of $\overline{V}_{{r} \hskip 0.5pt 0}$ = -6.1 $\pm$ 1.9 km s$^{-1}$ is statistically significant given the size of the 95\% CI. The $p$ value for the null hypothesis that $\overline{V}_{{r} \hskip 0.5pt 0}$ = 0 km s$^{-1}$ is 2.3 $\times$ 10$^{-3}$\%, which is small enough to rule out the null hypothesis (Ramsey \& Schafer 2012, Chapter 2, hereafter RS12). If this conclusion is wrong and $\overline{V}_{{r} \hskip 0.5pt 0}$ is actually closer to 0 km s$^{-1}$ by changing one of the other variables discussed in Section 5.3, there are still statistically significant differences from 0 km s$^{-1}$ in the radial profile of $V_{{r} \hskip 0.5pt 0}$ for E+R. Figure 23 shows an example of such results when $\phi_{\mbox{\footnotesize maj}}$ = 290\fdg1. The minimum of $V_{{r} \hskip 0.5pt 0}$  in Figure 23 is -3.3 $\pm$ 0.9 km s$^{-1}$.  The maximum is 4.6 $\pm$ 2.3 km s$^{-1}$. 

For a mean net radial velocity of $\overline{V}_{{r} \hskip 0.5pt 0}$ = -6.1 km s$^{-1}$, and an assumed distance of 5.1 Mpc from the mean of the estimates reported in the NASA/IPAC Extragalactic Database, the H{\footnotesize I} is transported inward across the 2.5 kpc oval distortion in 400 Myr, or 1.7 rotations of the oval distortion. There are many ways to account for what happens to this inflowing H{\footnotesize I}. Some of the H{\footnotesize I} is lost to photoionization and the formation of H$_2$, while some of it is created by radiative recombination of H{\footnotesize II} and the photodissociation of H$_2$. Some of the gas eventually becomes fuel for star formation, while some of it is swept up in feedback processes. In the presence of intense star formation, such as what is occurring in the inner pseudoring, galactic winds can entrain several $M_{\odot}$ yr$^{-1}$ of H{\footnotesize I}, and up to 10\% of that H{\footnotesize I} may escape into the intergalactic medium (Veilleux et al. 2005, and references therein).

The radial profiles of the column density, $N_{HI}$, and mass, $M_{HI}$, are consistent with an inflow of H{\footnotesize I} in the oval distortion. These are shown in Figure 24 for 10$^{\prime\prime}$ rings across the whole H{\footnotesize I} disk. Both profiles show a local minimum near the end of the oval distortion and increase with decreasing radius across the oval distortion. The radial profile of ${V}_{{r} \hskip 0.5pt 0}$ predicts that H{\footnotesize I} is piling up in a region at $\sim$ 140$^{\prime\prime}$. This corresponds to the peak in $M_{HI}$ within the region of the oval distortion. The calculation of $N_{HI}$ (e.g., Kwok 2007, Chapter 5) and conversion to $M_{HI}$ (e.g., Sparke \& Gallagher 2007, Chapter 5) is checked by summing all of the $M_{HI}$ in panel ($b$) of Figure 24 and comparing the result to the amount measured by W08. The result of 4.8 $\times$ 10$^8$ $M_{\odot}$ is in excellent agreement with W08 when adjusted for their adopted distance.  

From the sum of $M_{HI}$ in a ring and the results for $V_{{r} \hskip 0.5pt 0}$, one can estimate the mass flow rate, 
\begin{align}
\dot{M}_{HI} & \approx \int^{2\pi}_0 N_{HI} V_{r \hskip 0.5pt 0} \, \bar{r} \, d \theta ,\nonumber \\
& \approx M_{HI}\,V_{{r} \hskip 0.5pt 0}\,/\triangle r,
\end{align}
where $\bar{r}$ is the mean radius of a ring and $\triangle$$r$ is the ring width. The results are shown in Figure 25. The mean of $\dot{M}_{HI}$ is -0.25 $\pm$ 0.11 $M_{\odot}$ yr$^{-1}$. Excluding the result for $\dot{M}_{HI}$ near the end of the oval distortion that is indistinguishable from 0 $M_{\odot}$ yr$^{-1}$  decreases the mean to -0.30 $\pm$ 0.12 $M_{\odot}$ yr$^{-1}$. The largest result for $\dot{M}_{HI}$ is -0.53 $\pm$ 0.05 $M_{\odot}$ yr$^{-1}$. 

The mean of $\dot{M}_{HI}$ is similar to estimates of the star formation rate (SFR) reported in the literature. A summary of those estimates is provided in Table 5 for the whole disk and the inner pseudoring. They are adjusted for the adopted distance in this paper, ignoring small nonlinear corrections in the formulae used. The mean of the estimates for the whole disk is 0.65 $\pm$ 0.38 $M_{\odot}$ yr$^{-1}$, and that for the inner pseudoring is 0.28 $\pm$ 0.09 $M_{\odot}$ yr$^{-1}$. 

The similarities between $\dot{M}_{HI}$ and estimates of the SFR for the inner psuedoring are consistent with inward flows induced by the oval distortion playing a role in providing fuel for star formation in the nuclear region. For reasons pointed out in the second paragraph of this subsection, however, physically connecting these two processes is beyond the scope of this paper. Furthermore, not all of the gas is converted to stars in star-forming regions, and it is unknown what $\dot{M}_{HI}$ is in regions interior to the oval distortion. The most that can be inferred from the similarities between $\dot{M}_{HI}$ and estimates of the SFR is that the results for $V_{{r} \hskip 0.5pt 0}$ are reasonable. Determining the physical connections between {$\dot{M}_{HI}$} in oval distortions and the SFR in the nuclear regions of galaxies such as NGC 4736 therefore requires future applications of the method developed in this paper.

\begin{table*}[t!]
\centering
\caption{Estimates of the SFR}
\begin{tabular}{lllcl}
\hline
\hline
Region  & Method & Data & SFR & Reference \\
  & & & (M$_{\odot}$ yr$^{-1}$) & \\
\hline
Inner Pseudoring & CF   & H$\alpha$, 24$\micron$ & 0.19 & (1)\\
Inner Pseudoring  & CF   & NUV, 24$\micron$ & 0.32 & (1)\\
Inner Pseudoring & CF   & FUV, 24$\micron$ & 0.30 & (1)\\
Whole Disk & CF  & 1.4 GHz  & 0.59  & (2)\\
Whole Disk & SED  & IR - UV  & 0.56  & (3)\\
Whole Disk & CF  & FUV, 7.9$\micron$, 24$\micron$, 71$\micron$, 160$\micron$  &  0.84 & (4)\\
Whole Disk & CF  & H$\alpha$, 24$\micron$  &  0.26 & (4)\\
Whole Disk & CF  & H$\alpha$, 24$\micron$  & 0.39  & (5)\\
Whole Disk & SED  & IR - UV  & 1.28 & (6)\\
Inner Pseudoring & CF  & H$\alpha$ & 0.29 & (7)\\
\hline
\multicolumn{5}{l}{{\bf Notes}--Methods include conversion factors (CF) and fitting spectral energy distributions (SED).}\\
\multicolumn{5}{l}{{\bf References}--(1) van der Laan et al. (2015); (2) Heesen et al. (2014); (3) Lanz et al. (2013);}\\
\multicolumn{5}{l}{(4) Skibba et al. (2011); (5) Calzetti et al. (2010); (6) T09, (7) WB00.}

\end{tabular}
\end{table*}

\subsection{Future Application}

Future applications are also required to determine how robust this paper's method is for measuring $V_{{r} \hskip 0.5pt 0}$. Applying the method to simulated galaxies can constrain its usefulness for different $\epsilon$. Applications to other galaxies can investigate how the results compare to the properties of their host galaxies and to predictions from torque calculations (e.g., Garcia-Burillo et al. 1993, 2005, Quillen et al. 1995, Haan et al. 2009). Future applications will benefit from developing more sophisticated methods to determine $\phi_{\mbox{\footnotesize maj}}$. The dependence of the $\theta_o$ results for oval distortions on the coordinate system used needs further investigation. 

For galaxies without data suitable for the TW84 method, one can approximate $\Omega_o$ from the value of $\Omega$ near the end of the oval distortion. From the results for C+R, $\Omega$  = 0.68 km s$^{-1}$ arcsec$^{-1}$ at the end of the oval distortion. When this value is used for $\Omega_o$ in E+R, $\overline{V}_{{r} \hskip 0.5pt 0}$ = -6.4 $\pm$ 2.0 km s$^{-1}$, which is consistent with the adopted value given the size of the 95\% CIs. 

\section{Summary} \label{sec:sum}

This paper develops a method for measuring the net radial flow velocity of the gas in an oval distortion. It is applied to the H{\footnotesize I} in the oval distortion of NGC 4736. The findings are as follows:\\

\begin{enumerate}

\item The model describing the velocity field is linear in the unknown velocity components when $\alpha_{\mbox{\footnotesize kc}}$, $\delta_{\mbox{\footnotesize kc}}$, $\psi_{\mbox{\footnotesize inc}}$, $\phi_{\mbox{\footnotesize maj}}$, and $\theta_o$ are known. Of these, $\alpha_{\mbox{\footnotesize kc}}$, $\delta_{\mbox{\footnotesize kc}}$, and $\psi_{\mbox{\footnotesize inc}}$ are the most well known for NGC 4736. The method is the most sensitive to the accuracy of $\phi_{\mbox{\footnotesize maj}}$. 

\item The linear model describing the velocity field is degenerate in the unknown velocity components. The degeneracy is breakable using information about $\Omega_o$. 

\item The phase angle of the oval distortion is $\theta_o$ = 59\fdg8 $\pm$ 1\fdg4. This converts to a position angle of $\phi_o$ = 56\fdg8 $\pm$ 1\fdg6 in sky coordinates. The nuclear bar is offset from the oval distortion by 31\fdg9 $\pm$ 8\fdg4 in galaxy coordinates.

\item The angular frequency of the oval distortion is $\Omega_o$ = 0.67 $\pm$ 0.05 km s$^{-1}$ arcsec$^{-1}$. The oval distortion begins near the inner ultraharmonic (4:1) Lindblad resonance and  extends up to a corotation resonance. It is rotating more slowly than the nuclear bar. The angular frequency of the nuclear bar is 2.96 $\pm$ 0.36 km s$^{-1}$ arcsec$^{-1}$.

\item The H{\footnotesize I} is flowing inward at a mean rate of $\overline{V}_{{r} \hskip 0.5pt 0}$ = \linebreak-6.1 $\pm$ 1.9 km s$^{-1}$ in the region of the oval distortion. At this rate, it takes 400 Myr, or 1.7 rotations of the oval distortion, for the neutral hydrogen to travel the 2.5 kpc from the end to the beginning of the oval distortion.

\item The mean mass flow rate of the H{\footnotesize I} is $\dot{M}_{HI}$ = -0.25 $\pm$ 0.11 M$_\odot$ Gyr$^{-1}$. This is similar to the estimates of the SFR reported in the literature.

\end{enumerate}

\section{Acknowledgements} \label{sec:ack}

The authors acknowledge the helpful comments of the referee that greatly improved this paper. Dave Westpfahl is acknowledged for suggesting we work on the problem of measuring net radial velocities. Kathy Lynch is acknowledged for helping proofread this paper.  This research has made use of the NASA/IPAC Extragalactic Database (NED), which is operated by the Jet Propulsion Laboratory, California Institute of Technology, under contract with the National Aeronautics and Space Administration.

\clearpage
\appendix

\section{Model-fitting Methods} \label{subsec:meth_lsq}

\subsection{Least Squares}

This subsection of Appendix A provides a brief explanation of the model-fitting methods used. More details are available in Chapters 4 and 5 of A12.

The general, matrix form of a linear system of equations for $n$ data points is
\begin{equation}
{\boldsymbol d} = {\boldsymbol G}{\boldsymbol \beta}. 
\end{equation}
On the left-hand side of Equation (A.1), ${\boldsymbol d}$ is a column matrix of $n$ data points. This represents the left-hand sides of Equations (4), (5), (15), (18), and (25). On the right-hand side of Equation (A.1) is the matrix of independent variables, ${\boldsymbol G}$, and the column matrix of \underline{$o$} fitted variables, ${\boldsymbol \beta}$. The matrix ${\boldsymbol G}$ has $n$ rows and $o$ columns. The least-squares solution of Equation (A.1) minimizes SSR = $||{\boldsymbol d} - {\boldsymbol G}{\boldsymbol \beta}||^2_2$. 

Three different methods are used for minimizing the SSRs. The simplest method solves the normal equations,
\begin{equation}
{\boldsymbol \beta} = \{{\boldsymbol G}^T{\boldsymbol G}\}^{-1}{\boldsymbol G}^T {\boldsymbol d}. 
\end{equation}
In Equation (A.2) the superscript $T$ indicates matrix transpose. This method is used for fitting Equation (4),  E+R, Eonly, C+R, and Equation (18).

The results for when $\psi_{\mbox{\footnotesize inc}}$ and $\phi_{\mbox{\footnotesize maj}}$ are allowed to vary with radius are demonstrated using the Levenberg-Marquardt algorithm (Levenberg 1944, Marquardt 1963, Press et al. 1992) for nonlinear least squares. The algorithm starts with initial guesses for the fitted variables ${\boldsymbol \beta}$, and then calculates the Jacobian matrix of the model at row $i$ and column $j$,
\begin{equation}
J_{i,j} =  \frac{\partial f_i}{\partial \beta_j},
\end{equation}
where that $f$ is the nonlinear model, which for this paper is the right-hand side of Equation (4), and the subscript $i$ represents the $i$th data point, corresponding to $d_i$. The subscript $j$ represents the $j$th variable. A correction for the column matrix ${\boldsymbol \beta}$ is then found from calculating
\begin{equation}
\triangle {\boldsymbol \beta} = \{{\boldsymbol J}^T{\boldsymbol J} + \Lambda\, \mbox{diag}({\boldsymbol J}^T{\boldsymbol J})\}^{-1}{\boldsymbol J}^T \{{\boldsymbol d} - {\boldsymbol f}({\boldsymbol \beta})\},
\end{equation}
where that $\Lambda$ is a dampening factor. If there is an improvement in the SSRs, $\Lambda$ is decreased and the process is repeated. If not, it is increased until there is an improvement. The algorithm continues in this way until a desirable tolerance in $\triangle  \beta_j$ is achieved for all of the fitted variables.

First-order Tikhonov regularization is used for finding stable fits of Equation (25). This order of regularization penalizes oscillating solution instabilities while allowing for approximately linear gradients in the radial profile of the results. It minimizes $||{\boldsymbol d} - {\boldsymbol G}{\boldsymbol \beta}||^2_2 + \lambda^2||{\boldsymbol D}{\boldsymbol \beta}||^2_2$. The term added to the SSR includes the regularization parameter, $\lambda$, and the first-order difference operator, ${\boldsymbol D}$. 

Tikhonov regularization is implemented using singular value decomposition. Let
\begin{equation}
{\boldsymbol G} = {\boldsymbol U\boldsymbol S\boldsymbol V}^T,
\end{equation}
where ${\boldsymbol S}$ is a diagonal matrix of singular values, ${\boldsymbol U}$ is a basis vector spanning the data space, and ${\boldsymbol V}$ is a basis vector spanning the model variable space. Similarly for the difference operator,
\begin{equation}
{\boldsymbol D} = {\boldsymbol W\boldsymbol M\boldsymbol V}^T.
\end{equation}
The solution is then found by calculating
\begin{equation}
{\boldsymbol \beta} = {\boldsymbol V}^{-T} \{{\boldsymbol S}^{T} {\boldsymbol S} + \lambda^2\,{\boldsymbol M}^{T} {\boldsymbol M}\}^{-1} {\boldsymbol S}^{T} {\boldsymbol U}^{T} {\boldsymbol d}.
\end{equation}

The amount of regularization is determined by $\lambda$, which is chosen using the L-curve criterion (A12, Chapter 5). The L-curve criterion adopts the value of $\lambda$ that corresponds to the bottom left corner of the L shape in log-log plots of $||{\boldsymbol D}{\boldsymbol \beta}||_2$, the solution norm, as a function of $||{\boldsymbol d} - {\boldsymbol G}{\boldsymbol \beta}||_2$, the residual norm. This is a compromise between minimizing the SSRs, and the bias from minimizing $\lambda^2||{\boldsymbol D}{\boldsymbol \beta}||^2_2$. An example of an L-curve is provided in Section 4.3.

\subsection{Uncertainties}

The uncertainties are reported as 95\% confidence intervals (CIs). For plots, the entire interval is shown. For specific variables, they are reported as $\pm$ the half width (HW) of the 95\% CI. The HWs are calculated as,
\begin{equation}
\mbox{HW}=t\,\mbox{SE},
\end{equation}
where that $t$ is from the Student's $t$ distribution and $SE$ is the standard error (RS12, Chapter 2). The value of $t$ depends on the number of independent data points, the degrees of freedom, and the size of the CI. The jackknife method is used for estimating the SEs (Feigelson \& Babu 2012, Chapter 3), unless otherwise noted. The jackknife method estimates the SE from the variance of the results for $n$ solutions, each missing a different data point. The SEs for calculated variables (e.g., Equations (10) and (21)) are found by propagating the SEs for the fitted variables through the calculation. The SEs for the means of the previous estimates in Sections 5.3 and 5.4 are the standard deviations of the means divided by the square root of the number of estimates (RS12, Chapter 2).

This method for calculating the uncertainties assumes that the data are uncorrelated. The correlations are accounted for by dividing the number of data points, $n$, by the number of data points in the spatial resolution of the data, $n_c$, in the calculations of SE and $t$. For the velocity and $\theta_p$ models, the data are sampled in 10$^{\prime\prime}$ wide rings, so $n_c$ is the number of pixels in a correlated area. The correlated area for the H{\footnotesize I} data is approximated as the area in the FWHM of the synthesized beam. The region $r$ $<$ 10$^{\prime\prime}$ is therefore excluded when fitting velocity models to the H{\footnotesize I} $V_{\mbox{\footnotesize los}}$ data. The correlated area for the NIR data is approximated as the area in the FWHM of the point-spread function. For the $\Omega_p$ models, $n_c$ is the number of adjacent calculations of Equation (25), which is approximated as the number of pixels that fit in the major axis of the FHWM of the synthesized beam for the H{\footnotesize I} data. The value of $n_c$ is 47 for the velocity models, 4 for the $\theta_p$ model, and 7 for the $\Omega_p$ model.

\section{Experiments for $\theta_o$ Using Different Methods and Data}

Three experiments are performed that vary the data, methods, and coordinate system with the goal of better understanding the large difference between the adopted value of $\theta_o$ and the previous estimates. In the experiments involving unfiltered data, most of the foreground starlight from the Milky Way is removed using Source Extractor (BA96).

The first experiment evaluates how robust this paper's method is to different data.  Previous estimates used unfiltered $I$-band data (M95) and unfiltered 3.6$\micron$ data (Comer\'{o}n et al. 2014). The details of the data used by E04 for NGC 4736 are not provided by the author. 

The data in the first experiment include the H{\footnotesize I} data filtered for 180$^\circ$ symmetry, $S_{HI}$, an optical $I-$band image by Knapen et al. (2004) that is filtered for 180$^\circ$ symmetry, $S_{Iband}$, and the unfiltered $I_{3.6}$ data shown in panel ($a$) of Figure 1. The $\theta_o$ results for these data are 51\fdg5 $\pm$ 7\fdg1, 60\fdg4 $\pm$ 2\fdg1, and 56\fdg9 $\pm$ 2\fdg2, respectively, consistent with the adopted value given the size of the 95\% CIs. Plots of this experiment's results are not shown owing to their similarity to Figure 6, with the exception that the $\theta_p$ plots for the $S_{HI}$ and $I_{3.6}$ data show more scatter and larger 95\% CIs than the results shown in Figure 6 for the $S_{3.6}$ data. The conclusion of the first experiment is that different data are an unlikely explanation for the differences between the adopted value of $\theta_o$ and the previous estimates.

The second experiment fits ellipses to isophotes in sky and galaxy coordinates using the method of J87 to determine whether the results for this method depend on the coordinate system used. Both $S_{3.6}$ and $I_{3.6}$ for the NIR data are used for determining whether the results for this experiment depend on whether or not the data are filtered for 180$^\circ$ symmetry. Different initial values for the position angles in sky coordinates, $\phi_{oi}$, and phase angles in galaxy coordinates, $\theta_{oi}$, are used for determining whether the results for this experiment depend on these initial values. Initial values of $\epsilon_i$ = 0.23 are used for sky coordinates, and $\epsilon_i$ = 0.10 for galaxy coordinates. These $\epsilon_i$ are adopted from the mean values found by M95. The ellipse centers are assumed to be the kinematic centers adopted in this paper. The results are shown in Figure B26.

The results for sky coordinates shown in panel ($a$) of Figure B26 are in excellent agreement with the results shown in Figure 2 of M95. The mean of the  $\phi_o$ results for the $S_{3.6}$ data is 92\fdg7 $\pm$ 1\fdg9 when $\phi_{oi}$ = 93\fdg1 and 93\fdg1 $\pm$ 2\fdg2 when $\phi_{oi}$ = 56\fdg8. The mean of the  $\phi_o$ results for the $I_{3.6}$ data is 92\fdg5 $\pm$ 3\fdg5 when $\phi_{oi}$ = 93\fdg1. These are all consistent with the mean of the previous estimates given the size of the 95\% CIs.

The results for galaxy coordinates are shown in panel ($b$) of Figure B1. The mean of the $\theta_o$ results for the $S_{3.6}$ data is 53\fdg4 $\pm$ 5\fdg4 when $\theta_{oi}$ = 20\fdg6 and  54\fdg7 $\pm$ 6\fdg2 when $\theta_{oi}$ = 59\fdg8. The mean of the $\theta_o$ results for the $I_{3.6}$ data is $\theta_o$ = 55\fdg1 $\pm$ 4\fdg2 when $\theta_{oi}$ = 20\fdg6. These are all consistent with the adopted value of $\theta_o$ given the size of the 95\% CIs.

The $\epsilon$ results for the second experiment are inconsequential to the goal of the three experiments, so plots of them are not shown. The results for $\epsilon$ in sky coordinates consistently differ from the initial values by only a few percent and converge to a mean of $\approx$ 0.23. The results for $\epsilon$ in galaxy coordinates converge to a mean of 0.16 for the $S_{3.6}$ data when $\theta_{oi}$ = 20\fdg6, 0.17 for the $S_{3.6}$ data when $\theta_{oi}$ = 59\fdg8, and 0.14 for the $I_{3.6}$ data when $\theta_{oi}$ = 20\fdg6. 

There are two conclusions from the second experiment. The most important conclusion is that the results for fitting ellipses to isophotes in NGC 4736 depend on the coordinate system. The second conclusion is that the dependence on the coordinate system is not explained by the initial values for the location of the oval distortion or whether or not the data are filtered for 180$^\circ$ symmetry. 

The third experiment applies this paper's method in sky coordinates to determine whether the results for this paper's method also depend on the coordinate system. Similar to experiment 2, both $S_{3.6}$ and $I_{3.6}$ for the NIR data are used for determining whether the results for this experiment depend on whether or not the data are filtered for 180$^\circ$ symmetry. The results are shown in Figure B2. The mean of the $\phi_o$ results for the $S_{3.6}$ data is 92\fdg9 $\pm$ 2\fdg6, and that for the $I_{3.6}$ data is 93\fdg9 $\pm$ 3\fdg9.  The conclusion from the third experiment is that the dependence of the results on the coordinate system is independent of the method and whether or not the data are filtered for 180$^\circ$ symmetry.

\begin{figure*}[t!]
\center
\includegraphics[width=1\textwidth]{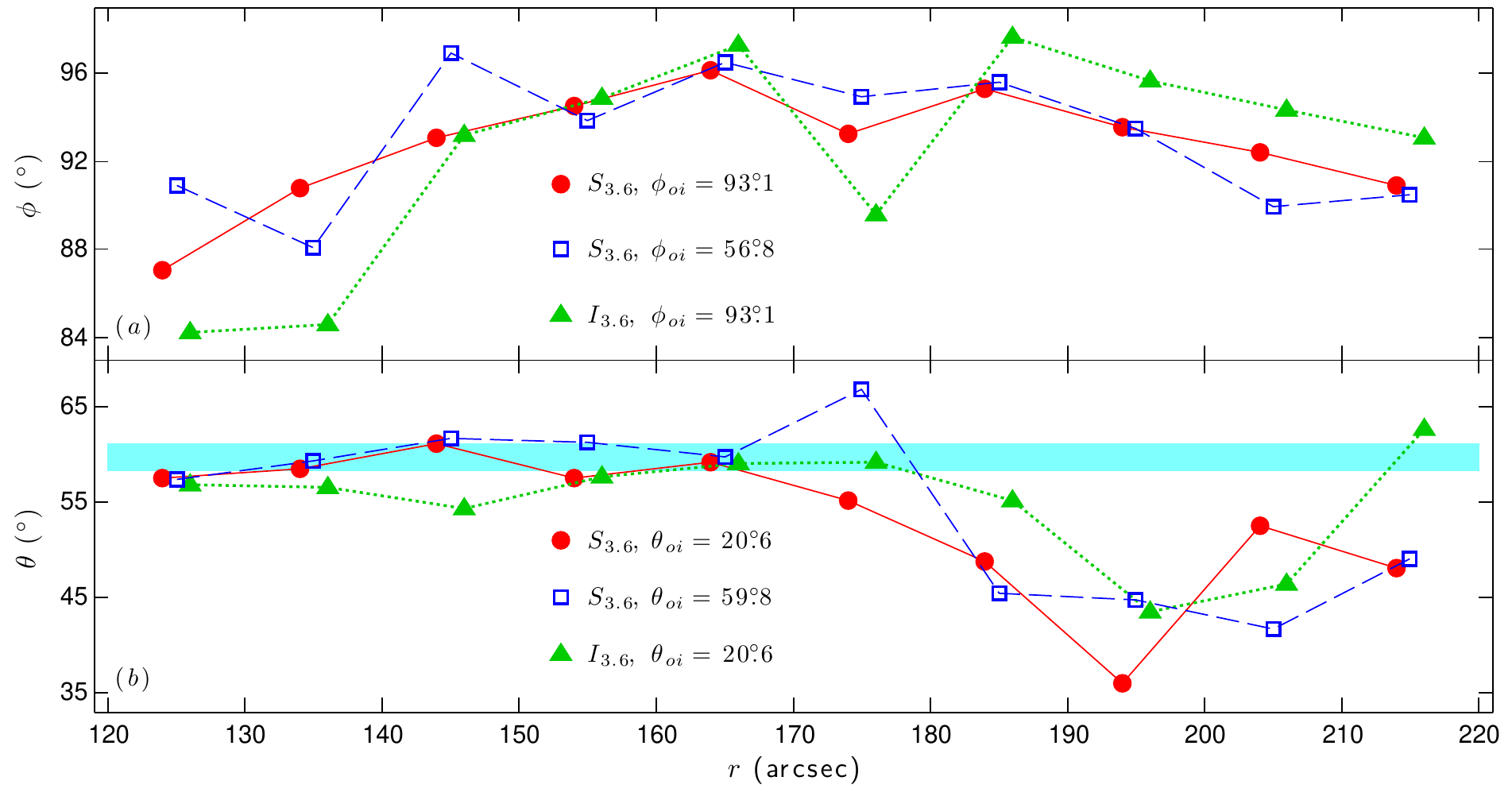} 
\caption{Results for ellipses fit to isophotes. Panel ($a$) shows the results for sky coordinates. Panel ($b$) shows the results for galaxy coordinates. The data type and initial values for the different markers are indicated in each panel. The red filled circles and green filled triangles are slightly offset from the center of each ring of data for clarity. The cyan shaded region in panel ($b$) shows the 95\% CI for the adopted value of $\theta_o$.}
\end{figure*}

\begin{figure*}[h!]
\includegraphics[width=1\textwidth]{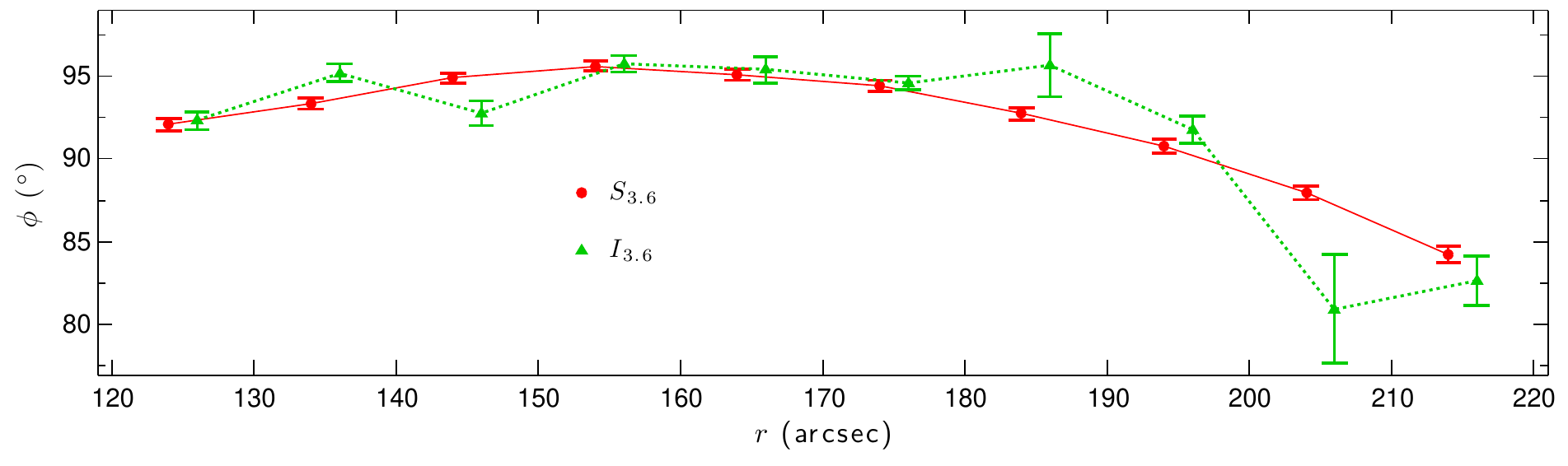} 
\caption{Results for calculations of Equation (4) from fits of Equation (11) in sky coordinates. The figure is formatted in a way that is similar to Figure B26 for ease of comparison. The data types for the different markers are indicated in the figure. The red filled circles and green filled triangles are slightly offset from the center of each ring of data for clarity. Note that the trend for $S_{3.6}$ is much smoother than the trend for $I_{3.6}$.}
\label{27}
\end{figure*}

\end{document}